\documentclass[floatfix,onecolumn,aps,prl,amsmath,amssymb]{revtex4}
\usepackage[utf8]{inputenc} 
\usepackage{caption}
\usepackage[dvips]{graphicx}
\usepackage{color}
\usepackage{amsmath,amssymb}
\begin{document}
\title{  Theory of High-Tc Superconductivity in Cuprates }
\author{E. C. Marino}\thanks{marino@if.ufrj.br}
\affiliation{Instituto de F\' isica, Universidade Federal do Rio de Janeiro, C.P. 68528, Rio de Janeiro, RJ, 21941-972, Brazil.}%
\date{\today}
\begin{abstract}

The essential physical processes underlying the phenomenon of High-Tc superconductivity in cuprates occur in the $CuO_2$ planes, found in these materials. The dynamics of the active electrons belonging to such planes is well described by the Three Bands Hubbard Model (3BHM). The complexity of such model, however, led the researchers to look for simpler and yet relevant alternatives.
In the attempts to circumvent the complexity of this model,two main simplified versions of the (3BHM) were considered. In the first alternative, one eliminates the doped holes and their respective sub-lattices by tying them to the $Cu^{++}$ electrons, thereby forming the so called Zhang-Rice singlets. The remaining dynamics consists in doping a Mott-Hubbard insulator and is described by the t-J Model. The second alternative
 maintains that the $Cu^{++}$ electrons form a square lattice of localized spins, while the doped holes move along the oxygen sub-lattices and undergo a Kondo like magnetic interaction with the localized spins, besides the Hubbard-like electric repulsion. This scenario is described by the Spin-Fermion-Hubbard Model. Most of the researchers in the field chose to follow the first road, while, I chose the second one. In this article I review in detail the reasons why that choice has led to a successful theory for High-Tc superconductivity in hole doped cuprates.

\end{abstract}

\maketitle

\begin{verse}{\em
... Somewhere ages and ages hence:\\

Two roads diverged in a wood, and I —\\

I took the one less traveled by,\\

And that has made all the difference.\\}
\end{verse}
\begin{verse}
Robert Frost - The Road not Taken 
   \end{verse}

\section{ 1) Introduction}

The discovery of high-Tc superconductivity in cuprates \cite{bm} was a landmark in physics. Yet, in the 40 years that have elapsed since then, the theories that have been proposed to describe the new phenomenon, were partially successful in describing the vast amount of experimental results that have been generated so far.

There is a general agreement that the simplest microscopic starting point for describing the High-Tc superconducting cuprates is Emery's Three Bands Hubbard Model \cite{3bhm,3bhm1,3bhm2,thesis_3band}. Considering the high degree of complexity of the 3BHM, however, simpler alternatives, yet capable to properly describe the High-Tc cuprates, have been sought.

A first possible approach for the simplification of the 3BHM is the t-J Model \cite{tj}, which is obtained from the former by means of a Schrieffer-Wolff transformation \cite{SW1,SW2,SW3}.  The so-called Zhang-Rice singlets \cite{zhangrice}, are formed by tying each doped hole, belonging to the Ligand molecular orbital of the oxygen lattices, to the copper electrons. The undoped system reduces to the  Heisenberg AF system and, as doping evolves, we have a doped Mott-Hubbard (Charge-Transfer) system.

A second approach to this problem has had the important contribution of J. Zaanen. He and co-workers obtained, out of the Three Bands Hubbard Model, an effective Hamiltonian describing the localized spins of $Cu^{++}$ ions and  the itinerant holes which are doped into the $O^{--}$ ions. This is known as the Spin-Fermion Hamiltonian \cite{SF1,sf,clust} and comprises an antiferromagnetic Heisenberg interaction term for the localized spins, which operates through the super-exchange mechanism, a hopping term for the holes and a Kondo-like magnetic interaction between the holes and the localized spins magnetic moments.
 
Most researchers have chosen the first alternative, based on the t-J Model. We have made the second choice and we will see that this choice is responsible, to a great extent, for the results of our approach to the cuprates.

In what follows I shall review the comprehensive microscopic theory, which we recently proposed for describing High-Tc superconductivity in cuprates as well as some of its main applications \cite{M1,M2,M3,M4,M5,M6}. This, besides being testable, has produced results that agree with a large number of different experimental data for several cuprate compounds.

In this paper, we exploit an important property of the lattice structure of the cuprates and describe it in Section 2.  This relates the presence or not of a dimerization in the underlying lattice to the nature of the the lowest energy states. For the Zhang-Rice singlets to be the lowest energy states produced by doping we need a non-dimerized lattice. Conversely, the occurrence of dimerization leads directly to a superconducting RVB-like state formed by a coherent sum of hole pairs such that each one of its two elements belongs to a different oxygen sub-lattice, as suggested by Anderson.
 
 In section 2, I review the derivation of the theory, which is basically, the Spin-Fermion-Hubbard model formulated on a lattice that contains three intertwined square lattices containing the copper and oxygen ions. Then, in section 3, I provide an overall review of different aspects of the complex phase diagram $T \times doping$ extracted from our theory. In section 4 I review the
 unified description of the resistivity in the different phases of the cuprates and in section 5 I review the effects that an applied external pressure has on the phase diagram of the cuprates. A comparative analysis between our theory and the ones associated to the t-J Model is presented at the conclusion, highlighting the reasons why the latter cannot be fully testable. 
 
 An Appendix with four sub-sections is provided, containing a pedagogical derivation of     the Heisenberg and Kondo terms of the Spin-Fermion Model by means of a Rayleigh-Schr\" odinger expansion using the pd-hybridization terms of the 3BHM. The material contained in this Appendix is already known in the literature (see \cite{dk}, for instance) and is added here for the benefit of the reader.

\section{2) Theory}
\subsection{ 2.1) Microscopic degrees of freedom}

The formulation of a theory, meant to describe any given physical system, necessarily starts from a judicious choice of the microscopic variables that will represent such a system at a microscopic level. In the case of the High-Tc cuprates, we assume  
 these are analogous to the Spin-Fermion \cite{SF1,sf} degrees of freedom, namely: a) localized, spin 1/2, magnetic moments located on a square lattice, associated to the copper ions; b) itinerant spin 1/2 holes introduced into the oxygen ions through doping \cite{nickelates,tranquada1,tranquada2}. The fact that the parent (undoped) compound is a Charge-Transfer insulator, pushes up the p-oxygen band, thus implying the doped holes go into the oxygen p-orbitals. The same does not seem to be true for nickelate compounds, for instance, where the parent compounds are Mott-Hubbard (not Charge Transfer) insulators. In this case the doped holes do not go necessarily to the oxygen ions \cite{nickelates}.

 The system resembles a Kondo lattice, which I recall, displays a network of localized spins associated to magnetic impurities introduced in a metal and itinerant electrons/holes, usually belonging to the system's conduction band and exhibiting a magnetic (Kondo) interaction with the localized spins. The resulting effective interaction among the magnetic impurities in the Kondo lattice is the so called RKKY (Rudermann, Kittel, Kasuya, Yosida) interaction. Conversely, the otherwise weakly interacting conduction band electrons acquire an effective interaction possessing a dynamically generated energy scale that separates a strongly interacting regime from a weakly interacting one.
 
 In the system of High-Tc cuprates, the localized spins are copper $3d^9$ spins, whereas the itinerant fermions are holes, introduced by doping in the oxygen p-orbitals, and moving on the intertwined bipartite lattice formed by the $p_x$ and $p_y$ orbitals of oxygen atoms. The very high Hubbard on-site repulsion energy, which is of the order of $9\ eV$, justifies treating the copper $3d^9$ electrons as localized spins.
 
  The localized copper spins undergo a mutual AF Heisenberg magnetic interaction, instead of the RKKY interaction, which is found in the Kondo lattice. Such AF Heisenberg interaction is provided by the super-exchange mechanism involving the hybridization of an oxygen p-orbital interconnecting two neighboring copper d-orbitals.

  Conversely, the effective interaction appearing among the doped holes, possesses, in addition to the usual Coulomb repulsion, a nearest neighbors attractive interaction between holes, that overcomes the former, thus producing the
``glue'' for the formation of Cooper pairs.

\subsection{ 2.2) Lattice}

\begin{figure}
	[h]
	\centerline{
\includegraphics[scale=0.8]{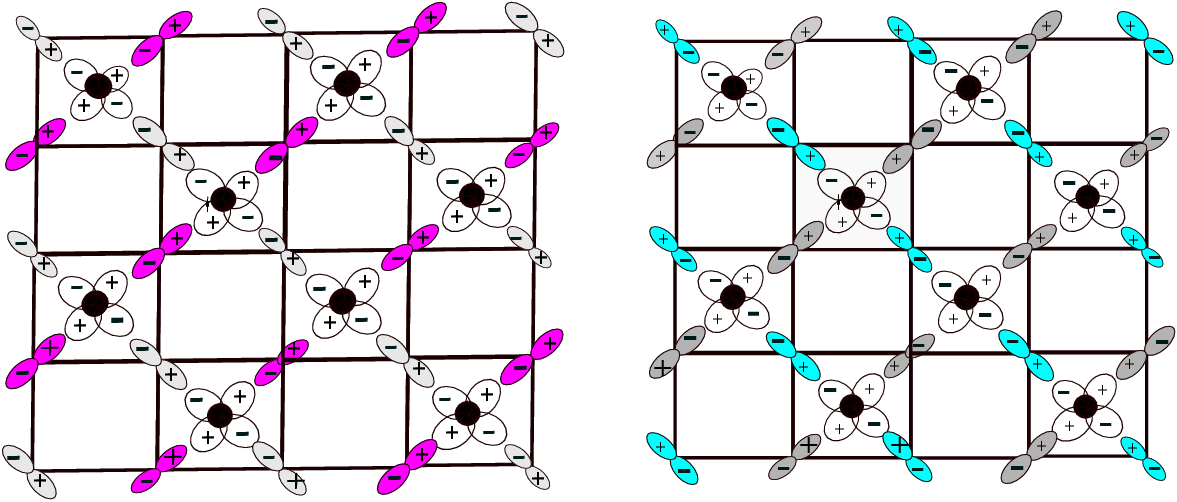}
	}
\caption{The $CuO_2$ lattice. The non-dimerized case is shown on the left, with oxygen sub-lattices $A$ and $B$ represented in magenta and gray.  The dimerized case is shown on the right, with oxygen sub-lattices $A$ and $B$ represented in cyan and gray. The copper ions are represented as black discs. }
	\label{ff1x}
\end{figure}

The stage where the dynamics responsible for High-Tc superconductivity in the cuprates unfolds is formed by $CuO_2$ layers, composed by three intertwined  square lattices: the copper square lattice with spacing $a$, and the bipartite oxygen lattices A and B with spacing $a/\sqrt{2}$, and the $p_x$ and  $p_y$ orbitals alternatively hybridizing with the copper $d_{x^2-y^2}$ orbitals. A peculiar feature of this structure is the possibility that the lobes of the oxygen p-orbitals organize themselves  in a dimerized form: ++ - - ++ - - ++ - - instead of + - + - + - + -. The two lattices are shown in Fig.\ref{ff1x}, the non-dimerized lattice in magenta/gray and the dimerized one in cyan/gray. 

It turns out that, in the presence of the copper ions in a square lattice, the dimerized  arrangement is energetically the most favorable. As we shall see, this fact is responsible for generating the effective  interaction between holes that leads to a SC phase in hole doped cuprates \cite{M1,M2,M3,M4,M5,M6}.

Interestingly, in the case of a dimerized oxygen lattice, a rotation of $90^\circ$ is equivalent to exchanging the A and B sublattices, namely: $A \leftrightarrow B$. We will see that this fact is ultimately responsible for the $d_{x^2-y^2}$ symmetry that is observed in the SC order parameter. 

\subsection{2.3)  Hole Doping:  Ligand Molecular Orbitals}

Now, let us consider what happens when holes are doped into the system and how the existence of a dimerized lattice influences the result.

Each copper atom is surrounded by four oxygen atoms, two in $p_x$ and two in $p_y$ orbitals. There is, by now, clear evidence that the doped holes go into the oxygen atoms, which were completely filled before doping. The introduced holes, for energetic reasons, preferably occupy,  molecular orbitals that strongly hybridize with the d-orbitals of copper, namely, Ligand orbitals. For this hybridization to happen effectively the  molecular orbital that hosts the holes must have the same symmetry as the copper d-orbital, namely, it must be $d_{x^2-y^2}$-symmetric.

 The doped hole, actually
goes into the state that is a linear combination of the appropriate Ligand $\psi_L$ and Anti-Ligand $\psi_{AL}$ molecular orbitals.The former is the only one that strongly hybridizes with the copper atom.

The specific form of the Ligand orbitals, is crucial for describing the physical properties of the system. What determines, ultimately, the form of the Ligand orbital is the symmetry of the corresponding d-orbital, to which the Ligand orbital couples. As we know, this has a $d_{x^2-y^2}$ symmetry, implying the Ligand orbital will be

\begin{eqnarray}
&\ &\psi_L=\frac{1}{2}\left [\eta_C\psi_{A}+ \eta_{C'}\psi_B 
\right ]
\nonumber \\
&\ &
\psi_L =\frac{1}{2} \left [ \eta_C\eta_A(1)\psi_{A}(1)+ \eta_C\eta_A(3)\psi_{A}(3)+ \eta_{C'}\eta_B(2)\psi_{B}(2) + \eta_{C'}\eta_B(4)\psi_{B}(4) \right ],
\nonumber \\
&\ &
\label{x7a}
\end{eqnarray}
where
\begin{eqnarray}
\psi_A=\frac{1}{\sqrt{2}}\left [ \eta_{A}(1)\psi_{A}(1)+ \eta_{A}(3) \psi_{A}(3) \right ],
\nonumber \\
\label{x7}
\end{eqnarray}
and
\begin{eqnarray}
\psi_B=\frac{1}{\sqrt{2}}\left [ \eta_{B}(2)\psi_{B}(2)+ \eta_{B}(4) \psi_{B}(4) \right ].
\nonumber \\
\label{x7}
\end{eqnarray}

The Anti-Ligand orbital, conversely, is given by the symmetric combination
\begin{eqnarray}
\psi_{AL}=\frac{1}{2}\left [  \psi_A(1) + \psi_A(3)+
\psi_B(2) + \psi_B(4)\right ].
\label{x7x}
\end{eqnarray}
 
In the previous expressions, $d^\dagger_{I;\alpha}$ creates an electron with spin $\alpha=\uparrow,\downarrow$ in the $3d^9$ orbital of copper located at $I$ and the $\psi^\dagger$-operators create holes in each of the four oxygen atoms, located at the vertices $A_1,B_2,A_3,B_4$ of a square, according to Fig. \ref{fi10}.
The sign factors $\eta_A,\eta_B=\pm 1$ originate in the overlap  integrals over the atomic orbitals and are determined by the sign of the lobes of the $p_x$ and $p_y$
oxygen orbitals, which overlap with the lobes of the $d_{x^2-y^2}$ copper orbitals, whose sign we denote by either $\eta_{C}=\pm 1$ or $\eta_{C}'=\pm 1$. 

\subsection{2.4)  Dimerization vs. Non-Dimerization}

The lattice associated to the $CuO_2$ planes in the cuprates admits two situations: either dimerized or not, as we explained in subsection 2.2. (see also \cite{M4,M5,M6}).

\subsubsection{Non-Dimerized}

\begin{figure}
	[h]
	\centerline{
		\includegraphics[scale=1.0]{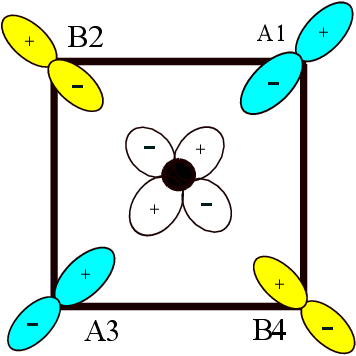}
	}
	\caption{The lattice configuration for the hole doping of the non-dimerized lattice. }
	\label{fi10}
\end{figure}

In this case the configuration (plaquette) representing one copper atom surrounded by the four nearest oxygen atoms is depicted in Fig. \ref{fi10}.

The Ligand molecular orbital, in this case, is given by
\begin{eqnarray}
\psi_L =\frac{1}{2} \left [- \psi_{A}(1)+ \psi_{A}(3)+\psi_{B}(2) - \psi_{B}(4) \right ].
\label{001}
\end{eqnarray}

The Ligand molecular orbital $\psi_L$ of oxygen, in the non-dimerized state has been used to construct the Zhang-Rice singlets  \cite{zhangrice}. This is made by combining in a zero total spin state (singlet), the spin $1/2$ electron located in the d-orbital of the copper atom with the spin $1/2$  doped hole, located in the Ligand molecular orbital  $\psi_L$, namely,
\begin{eqnarray}
\psi_{ZR}=\frac{1}{\sqrt{2}}\left [d_\uparrow \psi_{L\downarrow }- d_\downarrow \psi_{L\uparrow} \right].
\label{x62y}
\end{eqnarray}

Zhang-Rice (ZR) singlets have been playing a central role in the intense efforts that have been made in the pursuit of a theory describing superconductivity in cuprates. Since the Ligand molecular orbital is the lowest energy state, it follows that ZR singlets are energetically favorable states. Conversely, the fact that they combine degrees of freedom belonging to the three original bands of the cuprates, allows any description based on them, to use just a single band. This undoubtedly configures a substantial simplification in the formulation of a theory used to describe the cuprates. 

Nevertheless, we are going to provide arguments (see also \cite{M6} ) showing that a faithful description of the cuprates cannot forgo the use of the two oxygen square sublattices. 

\subsubsection{Dimerized Case}

\begin{figure}
	[h]
	\centerline{
		\includegraphics[scale=1.0]{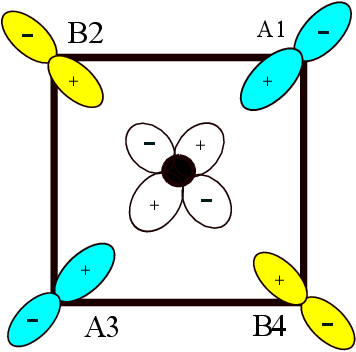}
	}
	\caption{Arrangement of $p_x$ (cyan ), $p_y$ (yellow) oxygen orbitals, which are closest to the localized $d$ (white) orbitals associated to an electron in a $Cu$ ion. }
	\label{fcucell}
\end{figure}

Lets assume now that we flip the p-orbitals located at oxygen atoms in the positions $A_1$ and $B_2$, according to Fig. \ref{fcucell}.

Now, the $\psi_A,\psi_B, \psi_L$ operators become  
\begin{eqnarray}
\psi_A =\frac{1}{\sqrt{2}}\left [  \psi_A(1) + \psi_A(3)\right ],
\nonumber \\
\psi_B =\frac{1}{\sqrt{2}}\left [  \psi_B(2) + \psi_B(4)\right ],
\nonumber \\
\psi_{L}=\frac{1}{\sqrt{2}}\left [  \psi_A - \psi_B \right ] = \frac{1}{2}\left [  \psi_A(1) + \psi_A(3)- \psi_B(2) - \psi_B(4)
\right ].
\label{x7aa}
\end{eqnarray}

We see the Ligand orbital is now completely different and possesses genuine $d_{x^2-y^2}$ symmetry. Consequently, the Zhang-Rice singlets are no longer energetically favored since the operator used for constructing these singlets (see (\ref{x62y}) ) is no longer the Ligand.

\subsection{ 2.4) Hamiltonian}

The starting point for deriving the Hamiltonian describing the relevant interaction of the system is the Three Bands Hubbard Model adapted for describing the interactions of the following actors: a) $d^\dagger_{I\sigma} $, creation operator of $Cu$ electrons, located at the site $I$; b) $p^\dagger_{A\sigma}$ and $p^\dagger_{B\sigma}$, creation operators of Oxygen p-electrons which live in the $A$ and $B$ sub-lattices; c) $\psi^\dagger_{A(B)\sigma}$, creation operator of oxygen holes on the $CuO_4$-plaquette that can be suitably expressed in terms of hole creation operators belonging to the $A$ and $B$ sub-lattices.

 The Hamiltonian that governs the dynamics of the system follows from the reasoning presented above, and is given by 

 \begin{eqnarray}
& &H=  E_d \sum_{I,\sigma} n^d_{I,\sigma}+ E_p \sum_{A,B,\sigma}\left[ n^p_{A,\sigma} + n^p_{B,\sigma} \right ] 
\nonumber \\
&\ &
+ U_p \sum_{A,B} \left[ n^p_{A,\uparrow}  n^p_{A,\downarrow}+ n^p_{B,\uparrow}n^p_{B,\downarrow} \right ] + U_d \sum_{I} n^d_{I,\uparrow} n^d_{I,\downarrow} 
+ U_{pd}\sum_{I,A,B} n^d_{I,\uparrow}\left[ n^p_{A,\downarrow} +  n^p_{B,\downarrow}
\right ]
\nonumber \\
&\ &
-\frac{t_{pd}}{2}\sum_{\langle I J \rangle}\sum_\sigma\left\{ \left[d^\dagger_{I,\sigma} p_{A\sigma} + p^\dagger_{A\sigma}d_{I,\sigma}
\right] +  \left[d^\dagger_{J,\sigma} p_{A\sigma} + p^\dagger_{A\sigma}d_{J,\sigma}
\right]+ \left[d^\dagger_{I,\sigma} p_{B\sigma} + p^\dagger_{B\sigma}d_{I,\sigma}
\right] +  \left[d^\dagger_{J,\sigma} p_{B\sigma} + p^\dagger_{B\sigma}d_{J,\sigma}
\right]  \right\}
\nonumber \\
&\ &
-\frac{t_{pd}}{2} \sum_{I}\sum_{\alpha} 
 \left \{d^\dagger_{I\alpha}
\left [\psi_{L }+\psi_{AL }\right ]_\alpha 
+ \left [\psi^\dagger_{L }+\psi^\dagger_{AL }\right ]_\beta d_{I\beta}\right\}
\nonumber \\
&\ & 
-t_p \sum_{A,B}\sum_\sigma
\left[ \psi^\dagger_{A\sigma} \psi_{B\sigma} +
\psi^\dagger_{B\sigma} \psi_{A\sigma} \right ] +U_p \sum_{A,B} \left[ n^A_\uparrow  n^A_\downarrow +  n^B_\uparrow  n^B_\downarrow   \right ],
\label{EEE}
  \end{eqnarray}
where
\begin{eqnarray}
n^d_{I,\sigma}= d^\dagger_{I,\sigma}d_{I,\sigma}\ \ ;\ \ 
n^p_{A,\sigma}= p^\dagger_{A,\sigma}p_{A,\sigma}\ \ ;\ \ 
n^A_\sigma= \psi^\dagger_{A\sigma} \psi_{A\sigma}
 \end{eqnarray}
 and $\sigma=\uparrow,\downarrow $. The corresponding $B$ operators are expressed in an obvious way $\psi_L$ and $\psi_{AL}$, are given, respectively, by (\ref{x7aa}) and  (\ref{x7x}).

 In the above expression the $U$-terms describe the on site Coulomb repulsion, whereas the $t$-terms describe the hopping of electrons/holes allowed by the hybridization of the corresponding orbitals.
 
  The effective Hamiltonian we use in the description of HTSC in cuprates has four terms, displayed in (\ref{hhh}).
Among these, we have the two last terms of (\ref{EEE}), namely
 $H_0$ and $H_U$. The first one describes the hole hopping between nearest neighbors, which belong, respectively, to the two sub-lattices $A$ and $B$. Then we have the second term, namely $H_U$, describing the on-site Hubbard Coulomb repulsive interaction for holes in each of the two sub-lattices $A$ and $B$.  The remaining two terms of our Hamiltonian are the Kondo-like magnetic interaction, between holes and localized spins, $H_K$ and the Heisenberg AF magnetic interaction, $H_{AF}$,  between neighbor localized  spins, $\mathbf{S}_I$. These two terms are obtained from the Hamiltonian (\ref{EEE}), respectively, as a 2nd. and 4th. order terms in an expansion in $t_{pd}$, as we show in the Appendix.
 
 Indeed, by  performing 2nd. order Rayleigh-Schr\" odinger perturbation theory, we demonstrate that the interaction between the doped itinerant holes and the localized copper spins is given by a Kondo interaction with Hamiltonian given by
\begin{eqnarray}
H_K =J_K\sum_{I,\textbf{R}_A\textbf{R}_B} \textbf{S}_I\cdot\mathcal{S}
\label{003a}
\end{eqnarray}
where
\begin{eqnarray}
J_K = t^2_{pd}\left[\frac{1}{\Delta E }+ \frac{1}{U_d-\Delta E}  \right ]
\label{jk00}
 \end{eqnarray}
 and $\Delta E = E_p-E_d$.
 The holes' spin is given by
\begin{eqnarray} 
\mathcal{S} =\sum_{\textbf{R}_A\in I}   \left [  \mathcal{S}_A(1) +
\mathcal{S}_A(3) \right ]- \sum_{\textbf{R}_B\in I}\left[
   \mathcal{S}_B(2) + \mathcal{S}_B(4) \right ]
   \end{eqnarray}
   
   \begin{eqnarray}
 \mathcal{S} =
\left[\sum_{\textbf{R}_A\in I}   \mathcal{S}_A -
\sum_{\textbf{R}_B\in I}     \mathcal{S}_B   \right ],
\label{abc2}
\end{eqnarray}
in the dimerized case and by 
 \begin{eqnarray}
 \mathcal{S} =
\sum_{\textbf{R}_A\in I}\sum_{\textbf{R}_B\in I}   \left [  \mathcal{S}_A(1) -
\mathcal{S}_B(2) - 
   \mathcal{S}_A(3) + \mathcal{S}_B(4) \right ]
   \label{abc3}
   \end{eqnarray}
in the non-dimerized state.

The factors $\eta_A,\eta_B,\eta_C,\eta_C'= \pm 1$ correspond to the signs of the lobes of the $p_x$, $p_y$
and $d_{x^2-y^2}$ orbitals, respectively.

As announced, the effective Hamiltonian we are going to use in order to describe the High-Tc cuprates is given by the four terms below:
\begin{eqnarray}& &
\hspace{-5mm}H_0 =-t_{p} \sum_{\textbf{R},\textbf{d}_i}\sum_{\sigma}\psi_{A,\sigma}^\dagger(\textbf{R})\psi_{B,\sigma}(\textbf{R}+\textbf{d}_i) +hc
\nonumber \\
&\ &
\hspace{-5mm}H_U =
U_p \sum_{\textbf{R}} n^A_\uparrow n^A_\downarrow + U_p \sum_{\textbf{R}+\textbf{d}_i} n^B_\uparrow n^B_\downarrow 
\nonumber \\
&\ &
\hspace{-5mm}H_{AF} = J_{AF}\sum_{\langle IJ \rangle} \textbf{S}_I\cdot  \textbf{S}_J
\nonumber \\
&\ &
\hspace{-5mm}
H_K =J_K\sum_{I} \textbf{S}_I\cdot\left[\sum_{\textbf{R}\in I} \eta_A \eta_C\  \mathcal{S}_A +
\sum_{\textbf{R}+\textbf{d}\in I} \eta_B \eta_C' \    \mathcal{S}_B   \right ],
\nonumber \\
&\ &
\label{hhh}
\end{eqnarray}
where
$\psi^\dagger_{A,\sigma}(\textbf{R})$, $\psi^\dagger_{B,\sigma}(\textbf{R}+\textbf{d})$ are the hole creation operators, with spin $\sigma=\uparrow,\downarrow$, on sites $\textbf{R}$ and $\textbf{R}+\textbf{d}$, respectively, of the $A,B$ oxygen sub-lattices. $n^{A(B)}_\sigma= \psi^\dagger_{A(B),\sigma}\psi_{A(B),\sigma}$ are the hole number operators for the sublattices $A,B$. Finally, $\textbf{S}_I$ are the spin operators of the localized $Cu^{++}$ ions. 

In the previous expressions
\begin{equation}
 J_{AF} = \frac{4t^4_{pd}}{\left (\Delta E + U_{pd} \right )^2}\left [ \frac{1}{U_d}+\frac{2}{2\Delta E +U_p} \right ],
 \label{jaf1}
\end{equation} 
is the effective Heisenberg coupling parameter generated by the super-exchange mechanism (see the Appendix, $\Delta E = E_p-E_d$ and $U_p,U_d,U_{pd}$ are the Hubbard Coulomb repulsion parameters ) 
and
\begin{eqnarray}
\mathcal{S}_{A(B)}=\frac{1}{2}\psi^\dagger_{A(B)\alpha} \vec{\sigma}_{\alpha\beta}\psi_{A(B)\beta} 
\label{hyhy}
\end{eqnarray}
 are the spin operator of holes belonging to the $p_x,p_y$ oxygen orbitals, which are associated, respectively, to the  $A$ and $B$ sub-lattices. 

These four terms comprise what is called the Spin-Fermion-Hubbard (SFH) model \cite{M1,M2,M3,M4,M5,M6}, which is obtained from the Spin-Fermion (SF) model \cite{SF1,sf} by the inclusion of the electric repulsion
term $H_U$. This Hamiltonian has been used in \cite{MM1} for studying some properties of the cuprates. The holes, however, were described as Dirac fermions in those studies.

The Spin-Fermion model preserves the 3 Bands structure from the original 3BHM, while the t-J model does not \cite{SF1}. This fundamental difference will prove to be crucial in the development of a model for High-Tc SC in cuprates. 

 We have used this Hamiltonian, Eq. (\ref{hhh}) in order to derive a substantial number of theoretical results that are in excellent agreement with the experimental data obtained from measurements made in several cuprate materials. \cite{M1,M2,M3,M4,M5,M6}.

\subsection{ 2.5)  Grand-Partition Function }
\begin{widetext}

The Grand-Partition function corresponding to the Hamiltonian (\ref{hhh}) is ($\beta=1/k_BT$)
\begin{eqnarray}
Z = \mathrm{Tr}_\psi\left\{ e^{-\beta\left[ H_0[\psi] + H_U[\psi] -\mu \mathcal{N} \right]}Z_{\textbf{S}_I}[\psi]\right\} 
\ \ ;\ \  
Z_{\textbf{S}_I}[\psi]=\mathrm{Tr}_{\textbf{S}_I}\left\{e^{-\beta\left[ H_{AF}[\textbf{S}_I] + H_K[\textbf{S}_I,\psi]\right ]}\right\}, 
\label{76b}
\end{eqnarray}
where $\mathcal{N}$ is the hole number operator.

The trace over the itinerant degrees of freedom (holes) can be made by means of a functional integral over the fermion fields, which are assembled in the form of a Nambu  fermion field \cite{M1,M2,M3,M4,M5,M6}, that is associated to the doped holes:
\begin{eqnarray}
\Psi_{a}=\left( \begin{array}{c}
\psi_{A,\uparrow,a} \\
\psi_{B,\uparrow,a}  \\
\psi^\dagger_{A,\downarrow,a} \\
\psi^\dagger_{B,\downarrow,a}
\end{array}               \right) \ \ \ ;\ \  \mathrm{Tr}_{\psi}= \int D\Psi D\Psi^\dagger,
\label{222y}
\end{eqnarray}

where $a=1,...,N$ is the number of planes per primitive unit cell.

   The trace over the localized degrees of freedom, conversely, can be
 conveniently made by using a base of coherent spin states \cite{ecm2,M1,M6}.
We can, then, express the trace over $\textbf{S}_I$ as a double functional integral on  $\textbf{n}$ and $\textbf{L}$ \cite{M1}, respectively the Antiferromagnetic and Ferromagnetic fluctuations of the localized copper spins:

   \begin{eqnarray}
&\ &\hspace{-5mm} \mathrm{Tr}_{\textbf{S}_I}= \int D\textbf{n}D\textbf{L}\delta(|\textbf{n}|^2-1)
\label{nl1}
\end{eqnarray}
The trace over the localized spin degrees of freedom, Eq.(\ref{76b}) 
can be written as \cite{M6}
\begin{eqnarray}
&\ &\hspace{-5mm} Z_{\textbf{S}_I}[\psi]=\mathrm{Tr}_{\textbf{S}_I}e^{-\beta H[\textbf{S}_I,\psi]}= 
\nonumber \\
&\ &\hspace{-5mm}\int  D\textbf{L} \exp\left\{-\int_0^\beta d\tau\left[ J_{AF}a^2 \sum_{I} |\textbf{L}_I|^2 + \frac{J_K}{2}\sum_{I}\sum_{i\in I}\textbf{L}_I\cdot \eta_i\mathcal{S}_{i}\right] \right \} \tilde Z_{NLSM},
\label{aa}
\end{eqnarray}
where $\tilde Z_{NLSM}$ is given by 
\begin{eqnarray}
 \tilde Z_{NLSM}=  \int D\textbf{n}\delta(|\textbf{n}|^2-1) \exp\left\{-\int_0^\beta d\tau\left[- J_{AF}s^2 \sum_{<IJ>}  \textbf{n}_I\cdot\textbf{n}_J +\sum_{<I>} +\frac{1}{c^2} \partial_\tau\textbf{n}_I\cdot \partial_\tau\textbf{n}_I \right. \right .\nonumber \\ \left . \left .
+J_K s \sum_{<IJ>}  \left [\textbf{n}_I+\textbf{n}_J\right ]\cdot \left[  \mathcal{S}_A -   \mathcal{S}_B  \right ]_{IJ}\right ] \right \},
\label{zn}
\end{eqnarray}
and
is the partition function of the Nonlinear Sigma Model \cite{M1,M2,M3,M4,M5,M6} augmented by the Kondo-like term, which describes the interaction between the itinerant spins and the antiferromagnetic fluctuations of the localized spins.

The quadratic functional integration over $\mathbf{L}$ in (\ref{aa}) can be made exactly yielding 
\begin{eqnarray}
Z_{\textbf{S}_I}[\psi] =  \exp \left \{-\beta H_1[\psi] \right\}\tilde Z_{NLSM},  
\label{9bx}
\end{eqnarray}

where
\begin{eqnarray}
&\ &
H_{1}[\psi]=
\nonumber \\
&\ &\frac{J^2_K}{8J_{AF}}\eta_A\eta_B\eta_C\eta_C'
\sum_{\textbf{R},\textbf{R}+\textbf{d}_i} 
\Big [\psi_{A\uparrow}^\dagger(\textbf{R})\psi_{B\downarrow}^\dagger(\textbf{R}+\textbf{d}_i)
 + \psi_{B\uparrow}^\dagger(\textbf{R}+\textbf{d}_i)
\psi^\dagger_{A\downarrow}(\textbf{R})\Big]
\Big [\psi_{B\downarrow}(\textbf{R}+\textbf{d}_i)\psi_{A\uparrow}(\textbf{R})
 + \psi_{A\downarrow}(\textbf{R}) \psi_{B\uparrow}(\textbf{R}+\textbf{d}_i) \Big].
\nonumber \\
& &\hspace{-3mm}
\label{h1aaa}
\end{eqnarray}

Now, making a 2nd order perturbative expansion in $t_p$ on $H_0 + H_U$, in (\ref{76b}) and using (\ref{9bx})
we obtain \cite{M1,M6} 
\begin{eqnarray}
Z = \int D\Psi D\Psi^\dagger \exp{\left\{-\beta\left [ H_0[\psi] + H_1[\psi] + H_2[\psi] -\mu \mathcal{N}\right ]\right\} }\ \tilde Z_{NLSM} 
\ \ \ ;\ \ \ Z=Z_{Holes}\tilde Z_{NLSM},
\label{9b}
\end{eqnarray}
where, 
\begin{eqnarray}
\hspace{-3mm}H_{2}[\psi]=
\hspace{1mm}-\frac{2t_p^2}{U_p}
\sum_{\textbf{R},\textbf{d}_i} \Big [\psi_{A\uparrow}^\dagger(\textbf{R})\psi_{B\uparrow}(\textbf{R}+\textbf{d}_i)
 + \psi_{A\downarrow}(\textbf{R}) \psi_{B\downarrow}^\dagger(\textbf{R}+\textbf{d}_i) 
\Big]
\hspace{1mm}
\Big [\psi_{B\uparrow}^\dagger(\textbf{R}+\textbf{d}_i)\psi_{A\uparrow}(\textbf{R})
+\psi_{B\downarrow}^\dagger(\textbf{R}+\textbf{d}_i) \psi_{A\downarrow}(\textbf{R})\Big].
\label{h2}
\end{eqnarray}


\subsection{ 2.6)  Effective Interactions of Holes }

Our theory produces two basic interactions among the itinerant doped holes: one attractive and one repulsive, described, respectively, by  $H_{1}[\psi]$ and
$H_{2}[\psi]$.

From Fig. 1 (see also\cite{M6}), we see that for nearest neighbors, which always belong to different sub-lattices $A,B$, we will always have
\begin{equation}
\eta_A\eta_C\eta_B\eta_C'=-1.
\label{1a}
\end{equation}
Consequently the effective interaction between holes described by $H_1[\psi]$,is always attractive for pairs of nearest neighbor holes. It originates from the magnetic mutual interaction between neighboring holes and localized copper spins. Observe that in the non-dimerized case the product 
$\eta_A\eta_C\eta_B\eta_C'$ is no longer exclusively negative. 

Indeed, we will have the product $\eta_A\eta_C\eta_B\eta_C'=-1$ for the pairs $A_1B_2$ and $A_3B_4$ whereas for the pairs $A_1B_4$ and $A_3B_2$, we will have $\eta_A\eta_C\eta_B\eta_C'=+1$. This unequivocally indicates: a) that the non-dimerized states have a higher energy than the dimerized one; b) the superconducting phase must occur in the dimerized state; c) Zhang-Rice singlets are not energetically favored in the dimerized (superconducting) state.

The $H_1[\psi]$ term of the effective holes' Hamiltonian produces an attractive interaction between nearest neighbor holes, which always belong to different oxygen sub-lattices. This term of the effective Hamiltonian is responsible for the formation of hole pairs, which, upon condensation, lead to the onset of a superconducting phase in cuprates. Each hole of the two sub-lattices, A and B is possibly surrounded by up to four holes, which belong to the other sublattice. 
We have showed in \cite{M6} that the ground-state of this system is an RVB-like \cite{RVB1,RVB2,BRVB,BRVB1,rk,rk1} state corresponding to the coherent linear combination of all possible dimer pairs formed by holes belonging, each other, to different sub-lattices. 

The term $H_2[\psi]$, conversely, 
 describes a repulsive interaction between neighboring holes, and consequently, the attractive interaction between neighboring electrons and holes, that is responsible for the formation of excitons.
 These are bound-states of a pair of neighboring electron and hole, which belong, each one, to different oxygen sub-lattices.
Excitons condense in the Pseudogap phase, and thereby generate all the Pseudogap phenomena, including the depletion of states near the Fermi level \cite{M1,M4}.

The strength of the effective interactions described, respectively, by  $H_{1}[\psi]$ and $H_{2}[\psi]$, according to (\ref{h1aaa}) and (\ref{h2}) is given by
           $$g_S=\frac{J^2_K}{8J_{AF}} \simeq 0.39 \ eV\  ; \ g_P =\frac{2t_p^2}{U_p} \simeq 0.30 \  eV,$$
for LSCO.
Here the parameters $J_{AF}$ and $J_{K}$ are given, respectively by (\ref{jaf1}) and (\ref{jk00}). 

For LSCO, the original 3BHM parameters are given by\cite{dp} :  $U_d=8.5 \ eV$, $U_p = 5.5 \ eV$, $U_{pd}=0.897\ eV$, $t_p=0.91 \ eV$, $t_{pd}= 1.48\ eV$, $\Delta_E=\epsilon_p-\epsilon_d=2.75\ eV$, which imply $J_K=1.17\ eV$ and $J_{AF}=0.43\ eV$.

Since $U_d > \Delta_E$, we see that the energy split between the two Hubbard bands is larger than the energy separation between the $d$ and $p$ orbitals, thus characterizing the undoped system as a Charge Transfer Insulator, with a gap          $\Delta_E$ \cite{cti1,cti2,cti3}.
\\
\bigskip

\subsection{ 2.7) The Superconducting and Pseudogap Order Parameters}

Applying a Hubbard-Stratonovitch transformation to the quartic fermion interactions in the Grand-Partition function, (\ref{9b})
we can rewrite our effective Hamiltonian in terms of the Hubbard-Stratonovitch fields 
$\Phi$ and $\chi$, as \cite{M1}

\begin{eqnarray}
 \hspace{-3mm}& \ & H_{eff}[\psi, \Phi, \chi]=-t_{p} \sum_{\textbf{R},\textbf{d}_i}\sum_{\sigma}\psi_{A,\sigma}^\dagger(\textbf{R})\psi_{B,\sigma}(\textbf{R}+\textbf{d}_i) +hc
\nonumber \\
\hspace{-3mm}&+&\frac{1}{g_S}\sum_{\textbf{R},\textbf{d}_i}\Phi^\dagger(\textbf{R},\textbf{d}_i) \Phi(\textbf{R},\textbf{d}_i)
+ \sum_{\textbf{R},\textbf{d}_i}\Phi(\textbf{R},\textbf{d}_i) 
\Big[\psi_{A\uparrow}^\dagger(\textbf{R})\psi_{B\downarrow}^\dagger(\textbf{R}+\textbf{d}_i)
+\psi^\dagger_{B\uparrow}(\textbf{R}+\textbf{d}_i) \psi_{A\downarrow}^\dagger(\textbf{R})
\Big]+ hc
\nonumber \\
&+&\frac{1}{g_P}\sum_{\textbf{R},\textbf{d}_i\in \textbf{R}}\chi^\dagger(\textbf{R},\textbf{d}_i)\chi(\textbf{R},\textbf+\sum_{\textbf{R},\textbf{d}_i}\chi(\textbf{R},\textbf{d}_i) \Big [\psi_{A\uparrow}^\dagger(\textbf{R})\psi_{B\uparrow}(\textbf{R}+\textbf{d}_i)
 + \psi_{A\downarrow}^\dagger(\textbf{R})\psi_{B\downarrow}(\textbf{R}+\textbf{d}_i)\Big]+ hc.
 \label{1aay}
\end{eqnarray}

 The Hubbard-Stratonovitch fields satisfy the field equations \cite{M1,M4}

\begin{eqnarray}
&\ &\Phi^\dagger(\textbf{R},\textbf{d}_i)= 
g_S\Big[\psi_{A\uparrow}^\dagger(\textbf{R})\psi_{B\downarrow}^\dagger(\textbf{R}+\textbf{d}_i)
+\psi^\dagger_{B\uparrow}(\textbf{R}+\textbf{d}_i) \psi_{A\downarrow}^\dagger(\textbf{R})
\Big]
\label{1yb}
\end{eqnarray}
and
\begin{eqnarray}
\chi^\dagger (\textbf{R},\textbf{d}_i) =  
g_P\Big [\psi_{A\uparrow}^\dagger(\textbf{R})\psi_{B\uparrow}(\textbf{R}+\textbf{d}_i)
 + \psi_{A\downarrow}^\dagger(\textbf{R})\psi_{B\downarrow}(\textbf{R}+\textbf{d}_i)\Big].
\label{1c}
\end{eqnarray}

Observe that $\Phi^\dagger$ is the creation operator of spin zero Hole Pairs  belonging to different sub-lattices, whereas $\chi^\dagger$
is the creation operator of spin one excitons formed by electrons and holes belonging to different sub-lattices.

The ground-state expectation value of these operators, namely, $\langle \Phi \rangle =\Delta$ 
and $\langle \chi \rangle = M$ are, respectively, order parameters for the SC and PG states of the system, the former being a Hole Pair condensate while the latter consists in an exciton condensate. They are given by \cite{M1,M4}. 
\begin{eqnarray}
\Delta(\textbf{k}) = \Delta\left[\cos k_+a' - \cos k_- a' \right]
\label{1bby}
\end{eqnarray}
and, also by
\begin{eqnarray}
M(\textbf{k})= M \left[\cos k_+a' - \cos k_- a' \right]
\label{1ccy}
\end{eqnarray}
where $k_\pm=\frac{k_x \pm k_y}{\sqrt{2}}$.

The SC and PG order parameters both have a d-wave symmetry, namely, change the sign under a $90^\circ$ rotation. This can be easily inferred from (\ref{1yb}) and (\ref{1c}) and by an inspection of Fig. \ref{ff1x} that shows such a rotation is equivalent to an exchange of sublattices: $A \leftrightarrow B$. 
Both these order parameters have nodal lines along the $\pm \hat{x}$ and $\pm \hat{y}$ directions.

Observe that the effective Hamiltonian (\ref{1aay}) possesses two continuum symmetries, namely, an overall $U(1)$ symmetry and a sublattice sensible "chiral" $U(1)$ symmetry. Indeed, the Hamiltonian possesses the following symmetries:

\begin{eqnarray}
\psi_{A\sigma} \rightarrow e^{i\theta }\psi_{A\sigma}
\nonumber \\
\psi_{B\sigma} \rightarrow e^{i\theta }\psi_{B\sigma}
\nonumber \\
\Phi \rightarrow e^{2i\theta }\Phi
\nonumber \\
\chi \rightarrow \chi,
\label{s1}
\end{eqnarray},
and a sublattice sensible "chiral" $U(1)$ symmetry 
\begin{eqnarray}
\psi_{A\sigma} \rightarrow e^{i\theta }\psi_{A\sigma}
\nonumber \\
\psi_{B\sigma} \rightarrow e^{-i\theta }\psi_{B\sigma}
\nonumber \\
\Phi \rightarrow \Phi
\nonumber \\
\chi \rightarrow e^{2i\theta }\chi,
\label{s1}
\end{eqnarray}
where $\theta \in \Re$.

These global continuum symmetries are spontaneously broken, respectively, in the  SC and PG phases of the cuprates, namely, $\Delta \neq 0, M =0$ in the former and  $\Delta =0 , M \neq 0$ in the latter. In \cite{M1,M6}, we carefully demonstrate that we cannot have both $\Delta \neq 0 $ and $ M
\neq 0$. The situation where both parameters vanish, 
$\Delta = 0, M =0$, conversely, corresponds to the "Strange Metal" phase. 

\subsection{ 2.6) The Thermodynamic Potentials}

The bridge between theory and experiments is made by the thermodynamic potentials. Let us explore then how to build such bridges.

We have
\begin{eqnarray}
Z=Z_{Holes} \tilde Z_{NLSM}.
\end{eqnarray}

The thermodynamic potential is introduced by 

\begin{eqnarray}
Z_{Holes}=\exp \{ -\beta \Omega_{ Holes} \}\ \ \ \ ; \ \ \ \tilde Z_{NLSM}=\exp \{ -\beta \Omega_{ Local} \} .
\label{effpot2a}
\end{eqnarray}

Consequently, the full thermodynamic potential is given by

\begin{eqnarray}
\Omega =\Omega_{ Holes} + \Omega_{ Local}.
\label{effpot3a}
\end{eqnarray}

Then, expanding the  Hubbard-Stratonovitch fields around their respective ground-state expectation values, namely 
\begin{eqnarray}
\Phi=\Delta +\phi \ \ \ ; \ \ \  \chi=M + \zeta,
\label{ax1}
\end{eqnarray}
and also
\begin{eqnarray}
Z_{Holes} \simeq Z[ \Delta,M,\mu ].
\label{ax2}
\end{eqnarray}
In \cite{M6} we present a complete analysis of the stability of this expansion and thereby carefully establish its validity.

At this point, a remark about the relation of our approach to the mean field approximation is in order.
Notice that we may write  the effective grand-partition function of the holes as 
 \begin{eqnarray}
Z_{holes} = \int D\psi D\bar \psi D\Phi D\chi \exp \left \{ -\beta H_{eff}[\psi, \Phi, \chi] \right \},
  \label{1y}
\end{eqnarray}
where $H_{eff}$ is given by (\ref{1aay}).

By applying (\ref{ax1}), (\ref{ax2}) to the equation above we obtain

\begin{eqnarray}
Z_{holes}(\Delta,M) = \int D\psi D\bar \psi  \exp \left \{ -\beta H_{eff}[\psi, \Delta,M] \right \}.
  \label{2y}
\end{eqnarray}
Notice that we perform a full functional integration over
the fermionic hole field and, consequently, the quantum fluctuations of this field are fully taken into account.

\subsection{ A) The Thermodynamic Potential of the Holes}

We now introduce a thermodynamic potential for the holes, $\Omega_{holes}=\Omega[\Delta,M,\mu]$, which is a function of the SC and PG order parameters, 
$\Delta $ and $M$, and of the chemical potential,
and out of these the partition functional is expressed as
\begin{eqnarray}
Z [ \Delta,M,\mu ] =\exp\Big \{ -\beta \Omega [ \Delta,M,\mu ]\Big \}. 
\label{effpot}
\end{eqnarray}
 In order to obtain the thermodynamic potential $\Omega =\Omega [ \Delta,M,\mu ]$, we write 
\begin{eqnarray}
&\ &Z_{holes }=\int D\Psi D\Psi^\dagger \exp\Big \{ \int d^2r\int_0^\beta d\tau\Big [ \frac{|\Delta|^2}{g_S} 
+\frac{|M|^2}{g_P} +N\mu d(x) \Big ]
+\Psi^\dagger \Big[ i \partial_\tau + \mathcal{H}[\Delta, M]-\mu \Big ] \Psi  \Big \},
\label{effpot}
\end{eqnarray}
where

\begin{eqnarray}
\mathcal{H}-\mu  =
\left(
\begin{array}{cccc}
-\mu & \epsilon + M & 0 &  \Delta
\\
\epsilon + M^* & -\mu & \Delta & 0
\\
0 & \Delta^{*} & \mu & -\epsilon -M^*
\\
\Delta^{*} & 0 & -\epsilon -M & \mu
\\
\end{array}
\right) 
\, .
\label{M2}
\end{eqnarray}

The fermion integration yields a determinant of $\mathcal{H}-\mu$ that can be expressed in terms of the eigenvalues of
this operator, namely,
\begin{eqnarray}
\varepsilon_\pm =\pm\sqrt{|\Delta|^2 +\left(\sqrt{\epsilon^2 + |M|^2}\pm \mu \right)^2}
\label{2xy}.
\end{eqnarray}

 Then, after performing the Matsubara summation over the frequencies associated to the $i\partial_\tau $ term of the action, we find the thermodynamic potential is given by \cite{M1}
 
\begin{eqnarray}
\Omega(\Delta, M, \mu)= \frac{|\Delta|^2}{g_S} +\frac{|M|^2}{g_P}+
N d(x) \mu  - 2TN \left(\frac{a}{2\pi} \right )^2 \int d^2k \left[
\ln \cosh\left(\frac{\varepsilon_+}{2T}\right)+
\ln \cosh\left(\frac{\varepsilon_-}{2T}\right)
\right ].
\label{1abc}
\end{eqnarray}

$g_s$ and $g_P$ are the coupling parameters of $H_1$ and $H_2$, which assume different values for each cuprate and  
$d(x)$ is a function (to be determined) of the stoichiometric doping parameter $x$. $\epsilon(k_x,k_y)$ is the tight-binding (kinetic) energy, namely, \cite{M1}
\begin{eqnarray}
\epsilon(k_x,k_y) = - 2t[\cos k_+ a' + \cos k_- a']
\label{3},
\end{eqnarray}\\
\bigskip
where $k_\pm=\frac{k_x\pm k_y}{\sqrt{2}}$ and $a'=\frac{a}{\sqrt{2}}$, with $a'$ and $a$ being, respectively, the lattice parameters of the oxygen and copper ions lattices.

The $d_{x^2-y^2}$ symmetric SC and PG order parameters are given, respectively, by (\ref{1bby}) and (\ref{1ccy}).
\end{widetext}


\subsection{ B) The Thermodynamic Potential of Localized Spins}

Let us rewrite the field associated with the antiferromagnetic fluctuations of the localized spins as
\begin{eqnarray}
\mathbf{n}_I=(\pi_1,\pi_2,\sigma)_I. 
\end{eqnarray}
Then, after integration over the  transverse components, $\pi_1,\pi_2 $, we find
the effective action  for the $\sigma$-component fields is given by
\begin{eqnarray}
&\ &Z_{Local }=\int D\sigma D\lambda \exp\{ -\beta\Omega\left [\sigma, \lambda\right ]\},
\label{effpot}
\end{eqnarray}

where the thermodynamic potential of the localized degrees of freedom is
 \begin{eqnarray}
\Omega\left [\sigma, \lambda\right ]=\frac{1}{2}\int_0^{\hbar\beta}d\tau\int d^2r\left[ \frac{1}{c^2} |\partial_\tau \sigma|^2+|\nabla\sigma|^2 + i\lambda\left[ \sigma^2 - \rho_0/3\right] \right ]
 -2 \mathrm{Tr} \ln \left[   \frac{1}{c^2} \partial_\tau ^2+\nabla^2 \right],
 \end{eqnarray}
where $\rho_0$ is the spin-stiffness and $\lambda$ is a Lagrange multiplier field enforcing the $\delta$-function present in.

\section{ 3) Phase Diagram}

\begin{figure}
	[p]
	\centerline{
		\includegraphics[scale=0.5]{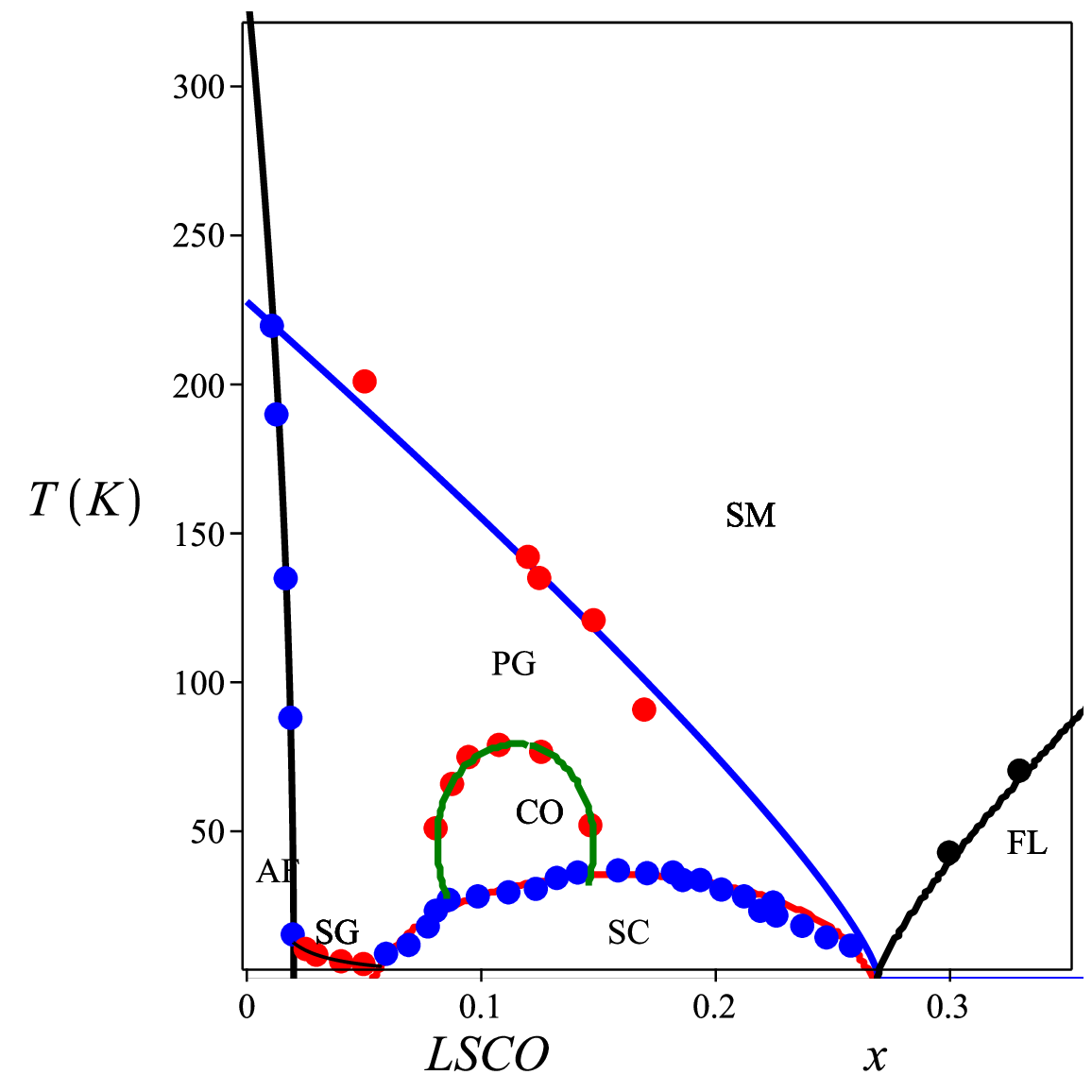}
	}
	\caption{The $LSCO$ $T \times x$ phase diagram \cite{M6}. The continuous lines correspond to analytic expressions for the transition temperatures provided by our theory for the high-Tc cuprates:  $T_N(x)$ (N\' eel-black),$ Tg(x) $(Spin-Glass-black), $Tc(x)$ (SC transition-red),$T_{CO}$ (CDW-Charge Ordering-green)
 $ T^*(x) $(PG transition-blue), $ T_{FL}(x); T_{FL}(x)= T^*(x-x_+)$ (FL transition-black)\cite{M1,M2,M3,M4,M5,M6}. Each of the curves displayed in Fig. \ref{f16xx} just contains a single adjustable fitting parameter.}
	\label{f16xx}
\end{figure}

Using the thermodynamic potentials found in the previous section, we may determine the full $T \times x$ phase diagram of the hole-doped cuprates \cite{M6}. We show below a sample of this procedure for the $T_c(x)$ and $T^*(x)$ lines of this phase diagram and refer the reader to \cite{M6} for the complete result, exhibiting $T_{Neel}(x)$,  $T_{SpinGlass}(x)$, $T_{ChargeOrder}(x)$, $T_{FermiLiquid}(x)$.
This complete phase diagram, for LSCO, is exhibited in Fig. \ref{f16xx} \cite{M6}.

\subsection{ 3.1) Superconducting, Pseudogap and Strange Metal Phases: The $T_c$ and $T^*$ transition lines }

We show below several applications of the theoretical results for the transition temperatures $T_c$ and $T^*$ in different cuprate materials.

The SC, PG and SM phases correspond, respectively, to $(\Delta \neq 0, M = 0)$
$(\Delta =  0, M \neq 0)$ and $(\Delta = 0, M = 0)$. 

\subsubsection{$\left [T_c(x)\right]$}

For the $T_c(x)$ line, we use the stationary condition 
\begin{eqnarray}
\frac{\partial \Omega}{\partial \Delta}=0\ \  ;\frac{\partial \Omega}{\partial M}=0\ \  ;\frac{\partial \Omega}{\partial \mu}=0,
\label{stcond}
\end{eqnarray}
 in the case 
\begin{eqnarray}
\Delta\neq 0\ ;M=0
\end{eqnarray}
and then taking the limit $\Delta \rightarrow 0$, we find the transition temperature \cite{M1}
$T_c$:
\begin{equation}
		\begin{cases}
			T_{c}(x) =\frac{\ln2 \,\ T_{max}}{\ln2 + \frac{\mu_C(x)}{2T_{c}(x)} - \frac{1}{2}\left(1-e^{-\frac{\mu_C(x)}{T_{c}(x)}}\right)},\hspace{0.5cm}x < x_{0}\\ \\
			T_c(x) =\frac{\ln2 \ \ T_{max}}{\ln\Big [1+ \exp\left[- \frac{\mu_C(x)}{T_c(x)} \right]  \Big ]},\hspace{1.6cm} x > x_{0}.\\
			\label{eqtc}
		\end{cases}
	\end{equation}

 The chemical potential, $\mu$, is a natural function of the amount of holes $y$ 
doped into the $CuO_2$ planes while the measured experimental data depend on the stoichiometric doping parameter $x$. It happens, however, that the precise relation between $x$ and $y$ is not known in general.
  
 Indeed, $\mu=\mu(y)$ but $y$, by its turn, is a function of the stoichiometric doping parameter $x$, namely, $y=f(x)$. $x$ is the parameter that controls the amount of doping \cite{M1} and the one, in terms of which, all the experimental consequences of doping are referred.
 
Given the fact that the function $y=f(x)$ is not known in general, it follows that Knowledge of $x$ does not imply, in general, knowledge of the chemical potential $\mu(x)$. 

In order to circumvent this obstacle,
we use the fact, implied by (\ref{eqtc}) \cite{M1,M2,M3,M4,M5,M6},
that the chemical potential along the curve $T_c(x)$, namely $\mu_C(x)$, vanishes at the optimal doping $x_0$ and directly write this as \cite{M1}
\begin{equation}
\mu_C(x)=2\gamma(x_0-x),
\label{mu}
\end{equation}
 where $\gamma$ is a parameter which must be determined for each compound. We, thereby trade the lack of knowledge about the fraction of the stoichiometric doping that actually goes into the  
 plane for our ignorance about the value of the parameter $\gamma$. This parameter, turns out to be the only adjustable parameter we use in the obtainment of the critical curve $T_c(x)$. 

    The maximum $T_c$, namely $T_{max}$ is given by \cite{M1}
	\begin{equation}
	    T_{max} =\frac{\Lambda \eta (Ng_S)}{2\ln 2} = \frac{\Lambda}{2\ln 2}\left [ 1-\frac{g_c}{N g_S}\right ] \simeq \frac{\Lambda}{2\ln 2}
        \exp\left\{-\frac{g_c}{N g_S}\right \},
	    \label{tmax}
	\end{equation}
where $\eta\left(N g_S\right)=1-\frac{g_c}{N g_S}$ and $g_S=\frac{J^2_K}{8J_{AF}}$.

$\Lambda=\frac{hv}{\xi}=0.018$ eV is a characteristic energy scale associated to the coherence length $\xi$, $N$ is the number of $CuO_2$ planes per unit cell and $g_c$ is the lowest threshold for the coupling parameter, namely $g_S > g_c=0.30\ eV$ \cite{M1,M2,M3,M4,M5,M6}. 

It is important to realize that the characteristic energy scale $\Lambda$ is not an adjustable parameter. It has a fixed value $\Lambda=0.018$ eV which is used for all cuprate compounds \cite{M1}.

Observe that for LSCO we use a symmetrized version of the equations to comply with the experimental observation that the SC dome is symmetrical for this compound.

Notice that the optimal transition temperature, given by (\ref{tmax}) can be viewed as the strong coupling expansion of the BCS transition temperature, given by the rhs of (\ref{tmax}).

The existence of a universality in the phase diagram of hole-doped cuprates has been reported in \cite{honma1}. This universality consists in the observation that, many hole-doped compounds have the same overall shaped phase diagram when expressed in terms of the variables $\tau_c\equiv T_c/T_{max}$ and $p=x/x_0$.
We can simply explain this result by using our expressions for $T_c(x)$ and $T_{max}$. Indeed, from 
	\begin{equation}
		\tau_c(p)=
	\begin{cases}
		\frac{\ln2}{\ln2 + \frac{\zeta(1-p)}{\tau_{c}(p)} + \frac{1}{2}\left(e^{-\frac{2\zeta(1-p)}{\tau_{c}(p)}}-1\right)}, p<1\\
		\frac{\ln2 }{\ln\Big [1+ \exp\left[ \frac{-2\zeta(1-p)}{\tau_c(p)} \right]  \Big ]}, p>1
	\end{cases},
	\end{equation}
where the dimensionless factor $\zeta$ is given by
	\begin{equation}
		\zeta=\frac{\gamma x_0}{T_{max}}.
	\end{equation}
    Since both $x_0$ and $T_{max}$
are experimentally determined for each cuprate compound, $\zeta$,
the single parameter that governs the phase diagram, is ultimately determined by $\gamma$.

\subsubsection{$\left [T^*(x)\right ]$}

Conversely, for the $T^*(x)$ line, we use

\begin{eqnarray}
\Delta=0\ ;M\neq 0.
\end{eqnarray}

\begin{widetext}

The stationary condition in the limit when $M\rightarrow 0$ leads to the following equation for the curve
separating the Pseudogap from the Strange Metal phases, namely, the Pseudogap Temperature:

\begin{equation}
		T^*(x) =\frac{\frac{\Lambda\tilde\eta(g_PN)}{2}}{\ln\Big [1+ \exp\left[- \frac{\tilde\mu(x)}{T^*(x)} \right]  \Big ]}.
		\label{eqtpg}
	\end{equation}
	
		In the above equation \cite{M1}
	\begin{equation}
\tilde\mu(x)=2\tilde\gamma (\tilde x_0 - x),
			\label{mutil}
	\end{equation}
	is the chemical potential along the curve $T^*(x)$.
		 $\tilde x_0$ is a parameter that will determine the point where $T^*\rightarrow 0 $ and is experimentally determined for each compound.
         
        The value of $\tilde\gamma$ must be adjusted for each compound \cite{M1,M4,M5} and is the only adjustable parameter for the curve $T^*(x)$. 
\begin{widetext}
\begin{figure}
	[h]
	\centerline{
\includegraphics[scale=0.9]{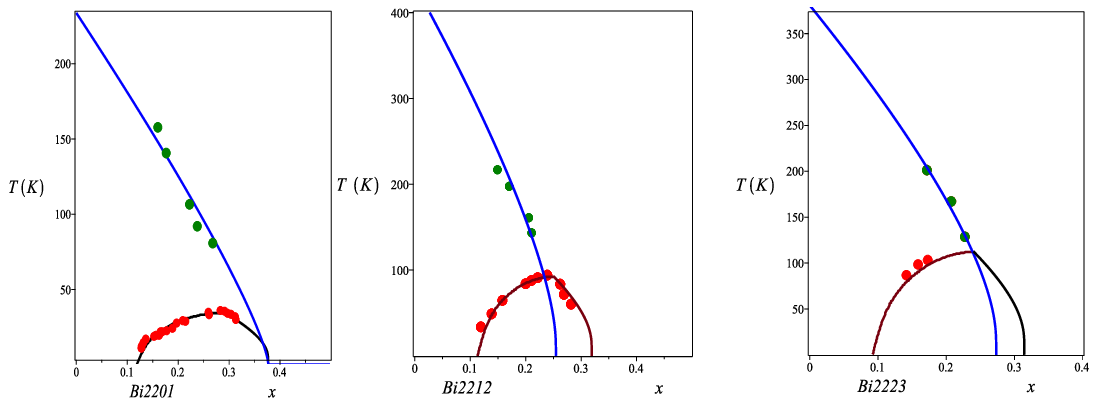}
	}
	\caption{The Lines $T_c(x)$ and T*(x), which determine the superconducting (SC)region and the line that separates the pseudogap (PG) phase from the strange metal (SM) phase for cuprates of the Bi-Family. Experimental data from \cite{bis1,bis2,pgBi,b4}.    }
	\label{ff5}
\end{figure}
\end{widetext}

\begin{figure}
	[h]
	\centerline{
		\includegraphics[scale=1.0]{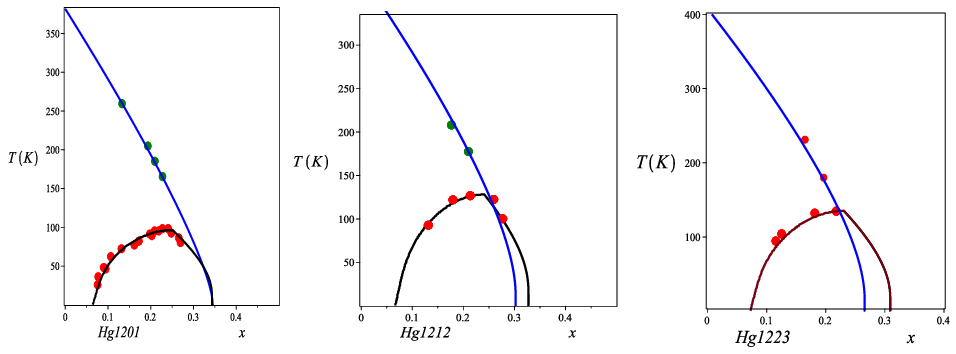}
	}
	\caption{The Lines $T_c(x)$ and T*(x), which determine the superconducting (SC)region and the line that separates the pseudogap (PG) phase from the strange metal (SM) phase for cuprates of the Hg-Family. Experimental data from \cite{mer1,7}. }
	\label{ff55}
\end{figure}

\begin{figure}
	[h]
	\centerline{
		\includegraphics[scale=0.3]{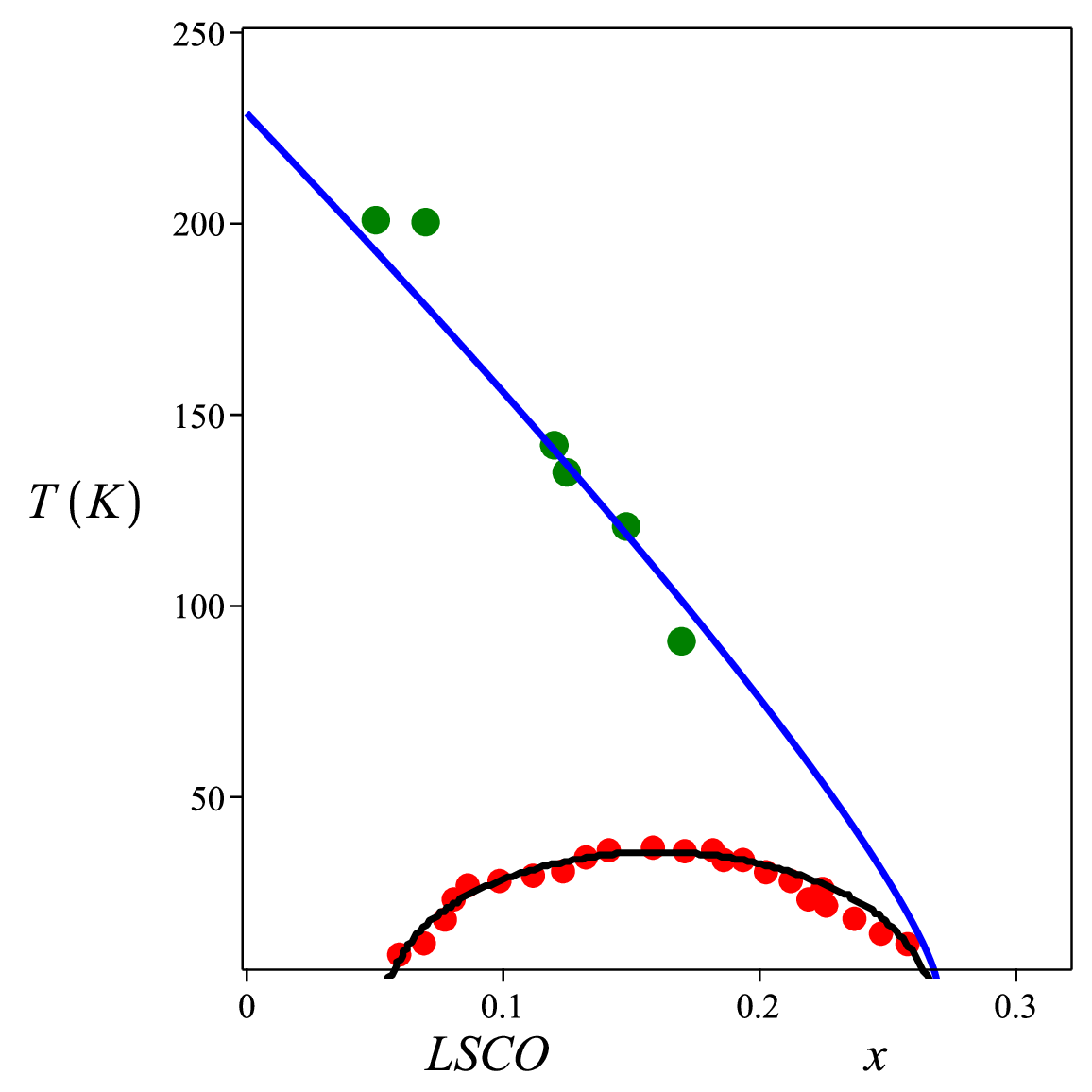}
	}
	\caption{The Lines $T_c(x)$ and T*(x), which determine the superconducting (SC)region and the line that separates the pseudogap (PG) phase from the strange metal (SM) phase for LSCO. Experimental data from \cite{l1,l2,003}.}
	 	\label{ff8}
\end{figure}

\begin{table}[!ht]
\begin{tabular}{|c|c|c|c|c|c|c|c|c|}
\hline\hline
           & N & $\zeta$  & $T_{max} (eV)$ &  $x_0$   & $\gamma$ (eV) & $\eta(N)$    & $\eta(1)^{1/N}$ &  $\gamma x_0 \eta(1)^{1/N}$\\ \hline \hline
Bi2201 & 1 &   1.16   & 0.0030 &   0.29   &      0.012    & 0.23077      & 0.23077 & 0.0008030      \\ 
Bi2212  & 2 & 1.256    & 0.0080 &  0.25    &    0.0389     & 0.61538      & 0.48038 & 0.004826     \\
Bi2223   & 3 & 1.117   & 0.0097 &   0.24  &    0.049      & 0.74358      & 0.61337 & 0.00664580   \\ \hline
Hg1201 & 1 &  0.928    & 0.00835 & 0.25     &    0.031      & 0.61577      & 0.61577 & 0.004772           \\
Hg1212 & 2 &  0.951    & 0.0111 &  0.24    &      0.044    & 0.80788      & 0.78471 &  0.008286          \\ 
Hg1223 & 3 &  1.005    & 0.0117 &  0.23   &     0.052     & 0.87192       & 0.85076 & 0.01           \\ \hline
LSCO    & 1 &  1.032   & 0.0031 &   0.16  &   0.020       &   0.23870    & 0.23870 & 0.00076     \\ \hline
\end{tabular}
\caption{The parameters used for obtaining the $T_c(x)$ curves. $T_{max}$ and $x_0$ are fixed experimental inputs and $\eta$ is determined by the former. Only $\gamma$ has been adjusted, in order to fit the experimental data. The last column displays the value obtained for the combination $\gamma x_0 \eta(1)^{1/N}= \zeta T_{max} \eta(1)^{1/N}$.}
\label{t1}
\end{table}

\end{widetext}

Notice that the results depicted in Figs. 1, 2 suggest that for single layered materials such as Hg1201 and Bi2201 the PG transition line goes to zero asymptotically in the same point as the SC line, whereas for multiple layered materials it goes to zero inside the SC dome.

We show below the parameters relevant for the obtainment of the $T^*(x)$ curves
for different cuprates

\begin{table}[!ht]
\begin{tabular}{|c|c|c|c|c|c|c|}
\hline\hline
           & N   &   $\tilde{x}_0$   & $\tilde{\gamma}$ (eV) & $\tilde{\eta}(N)$    & $\tilde{\eta}(1)^{1/N}$ &  $\tilde{\gamma}\tilde{x}_0 \tilde{\eta}(1)^{1/N}$\\ \hline \hline
Bi2201 & 1       &    0.376   &      0.132    & 0.01618      & 0.01618 & 0.0008030       \\ 
Bi2212  & 2      &   0.24    &    0.158     &  0.50809             & 0.12720 & 0.0048234     \\
Bi2223   & 3     &    0.245  &    0.1164      & 0.67205               & 0.25292 & 0.0066423    \\ \hline
Hg1201 & 1       &  0.343     &    0.186      & 0.07480      & 0.07480 & 0.004772            \\
Hg1212 & 2       &   0.28    &      0.1082    & 0.53740              & 0.27349 &  0.008286          \\ 
Hg1223 & 3       &   0.24  &     0.118    &   0.69159          & 0.42134 & 0.01          \\ \hline
LSCO    & 1      &     0.269  &   0.180       &   0.01565    & 0.01565 & 0.00076       \\ \hline
\end{tabular}
\caption{The parameters used for obtaining the $T^*(x)$ curves. Only $\tilde\gamma$ has been adjusted in order to fit the experimental data. $\tilde{x}_0$ is a measured parameter, and has a fixed determined value for each compound. The last column displays the value obtained for the combination $\tilde{\gamma}\tilde{x}_0 \tilde{\eta}(1)^{1/N}$. Notice that it is identical to $\gamma x_0 \eta(1)^{1/N}$.}
\label{t2}
\end{table}

\subsection{ 3.2) N\'eel Phase }

We have seen in \cite{M6} that, as we dope the system by the introduction of holes  in the oxygen sites, the N\'eel temperature will quickly decrease. This happens because there is a threshold spin stiffness, which decreases with doping, beyond which $T_N(x)$ vanishes.

We show in \cite{M6} that the N\'eel temperature for a given doping parameter $x$, namely $T_N(x)$ satisfies
\begin{eqnarray}
		T_N(x)=\frac{\rho_s(x)}{\left |\ln \left[1 - e^{-\rho_s(x)/T_N(x)}\right] 
        \right |},
 \label{n4}
\end{eqnarray}
where
\begin{equation}
   \rho_s(x) = \rho_s \left (1 - \frac{x}{x_{AF}} \right )^{1/2}
    \label{10},
\end{equation}
and $ \rho_s(0) = \rho_s$ is the spin stiffness.

From (\ref{n4}) we obtain the N\'eel temperature at zero doping: 
\begin{equation}
   T_N(x=0) = \frac{\rho_s}{\ln 2}
    \label{10aa}.
\end{equation}

For LSCO, we have \cite{M6}: $x_{AF}=0.020 $eV and $\rho_s= 0.0194$ eV


\subsection{ 3.3) The Spin-Glass Phase: Tg (x)}
\begin{widetext}

When we dope a hole into the  $CuO_2$ lattice \cite{M6}, this hole modifies the AF coupling between adjacent localized copper spins according to the different locations into which the  the doped holes may go, namely, 
\begin{eqnarray}
    J_{AF} \longrightarrow \left \{ J_{AF}+ J_0, J_{AF}-J_0\right \},
\end{eqnarray}
where
 $$
J_0= \frac{4J_K^2}{J_{AF}}.
  $$
The realization of each possibility
depends on whether the hole goes into either the A or B oxygen sub-lattices.
In the case where no doped holes go into the oxygen atoms connecting these spins, 
the resulting coupling would just remain $J_{AF}$.
It turns out that, we have
a stochastic system 
where the local couplings $J_{IJ}$ are random and
given by
 \\
 $J_{IJ}= \begin{cases}
\ \ \ \ \ J_{AF}	\\
\ \ \ \ \ J_{AF} + J_0	\\
\ \ \ \ \ \ J_{AF}-J_0	\\
	\end{cases}$,
 with the associated probabilities given by
   
$\begin{cases}
\ \ \ \ \ \  P[J_{AF}]=1-x	\\
\ \ \ \ \ \ P[J_{AF} + J_0]= \frac{x}{2}	\\
\ \ \ \ \ \ P[J_{AF}-J_0] = \frac{x}{2}	\\
			\end{cases}
   $
where $x$ is the doping parameter.

 \begin{figure}
	[h]
	\centerline{
		\includegraphics[scale=0.45]{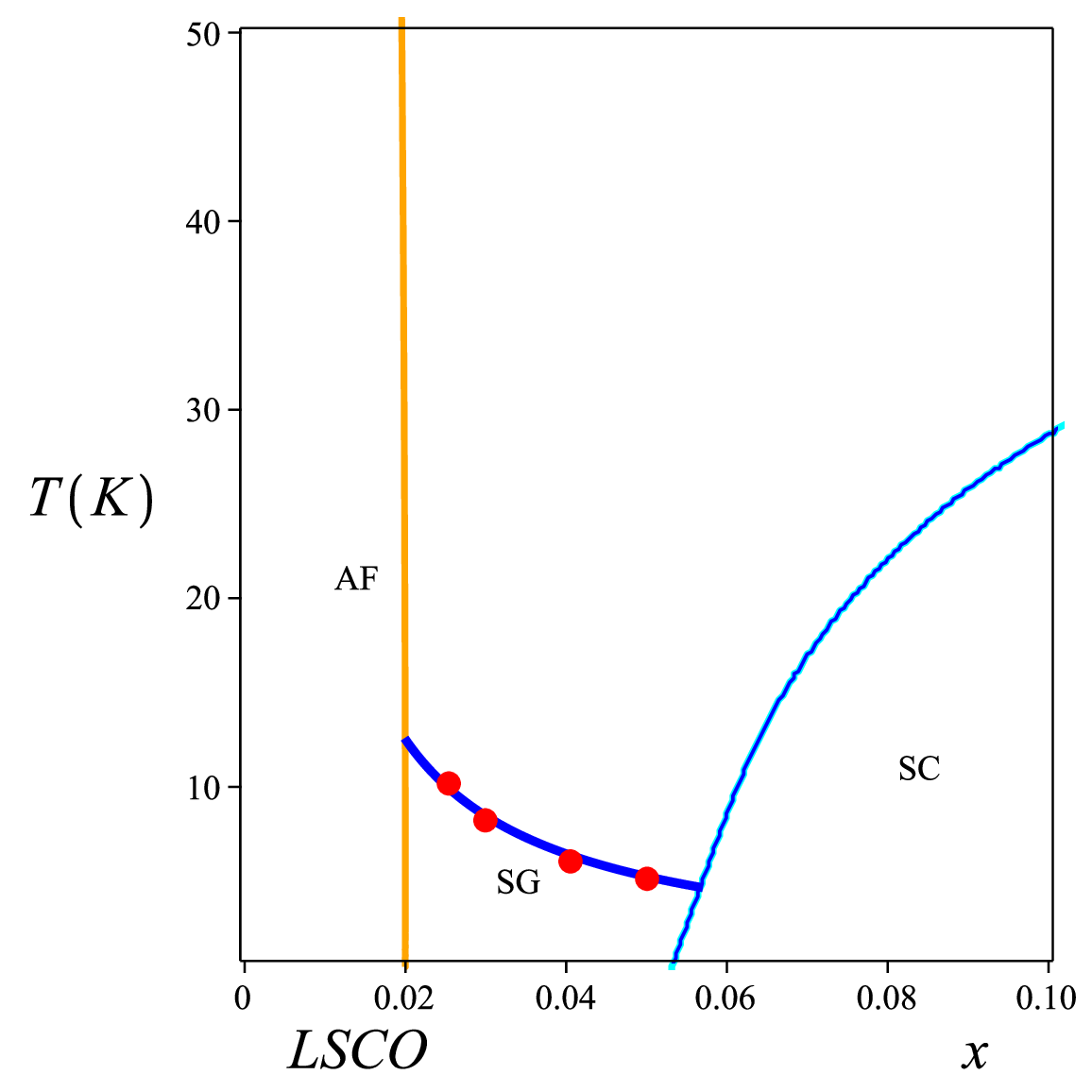}
	}
	\caption{The $LSCO$ $T \times x$ phase diagram (partial view). The continuous lines correspond to analytic expressions provided by our theory for the high-Tc cuprates for: a) the N\'eel transition line (orange); b) for the Spin-Glass transition, line (blue), c) for the Superconducting transition line (cyan)
 \cite{M1,M2,M3,M4,M5}. The experimental data for the Spin-Glass transition (red circles) are taken from \cite{nieder}.}
\label{f8}
\end{figure}

The average coupling  and variance will be given, respectively, by 
\begin{eqnarray}
& &
\langle J_{IJ}\rangle = J_{AF},
\nonumber \\
& &
\Delta J_{IJ}=  \sqrt{x} J_0.
\label{vv}
\end{eqnarray}

The present system coincides with the one obtained from a previous analysis of the Heisenberg antiferromagnet with a random coupling on a square lattice \cite{ecm2,sg1,sg2,sg3}. The results obtained in those studies do apply here as well.

A particularly interesting result was the obtainment of the transition line $T_g(x)$, delimiting the SG phase, which is characterized by infinite on-site time correlations that are associated to a nonzero Edwards-Anderson SG order parameter.
This is given by \cite{ecm2,sg1,sg2,sg3} 
\begin{eqnarray}
	 T_g=\lambda \frac {\varphi^{3/4}}{ \ln \left[   \frac{\Lambda_a}{T_g}\right ]-\frac{1}{2}\ln [ 1 + \varphi ] },
 \label{a}
\end{eqnarray}
where
$\Lambda_a=\frac{hc}{a}\simeq 0.1843$ eV and
\begin{eqnarray}
\varphi = \frac{3\pi}{x}  \left( \frac{J_{AF}}{2 J_K }\right )^4.
\end{eqnarray}
For LSCO, we have $\lambda=0.0085$,  and
$$
\varphi=\left ( \frac{0.0107}{x}\right ).
$$
The comparison of the results with the experimental data are excellent, as we can see from Fig. 8.

\end{widetext}

\subsection{ 3.4) Charge Order and the Fermi Surface Formation (Fermi Pockets)}

The Fermi surface can be defined as the manifold for which the eigenvalues of $\mathcal{H}-\mu $ are equal to zero, where $\mu(x)$ is the chemical potential of the holes.
Imposing the condition that such eigenvalues, given by (\ref{2xy}), vanish, we obtain the following equation for the Fermi surface points, (for $\Delta=0$)
\begin{eqnarray}
\mu(x) =\pm \sqrt{\epsilon(\mathbf{k})^2 + M(\mathbf{k})^2 },
\label{cd1}
\end{eqnarray}
 where $\epsilon(\mathbf{k})^2 = V^2\left( \cos k_1a + \cos k_2a\right)^2$ is the tight-binding energy eigenvalue and  $M^2 \left( \cos k_1a - \cos k_2a\right)^2$ is the DDW Pseudogap order parameter \cite{M1}.

We may determine the occurrence of Fermi pockets explicitly, by expanding the functions, $\epsilon(\mathbf{k})^2 $ 
 and  $M(\mathbf{k})^2$
around the points $(k_1,k_2)= (\pm \frac{\pi}{2a},\pm  \frac{\pi}{2a})$. These are four ellipses centered at such points, with semi-axes given, respectively, by (in $rlu$, reciprocal lattice units)

\begin{eqnarray}
A=\frac{\mu(x)}{2\pi\sqrt{2}V} \ \ \  ; \ \ \ B= \frac{\mu(x)}{2\pi\sqrt{2}M}.
\label{AB}
 \end{eqnarray}
where $V=\hbar c/a$.

Now applying the stationary condition 
\begin{eqnarray}
\frac{\delta \Omega}{\delta M(\mathbf{k})} = 0 ,
\end{eqnarray}
we obtain, 
\begin{eqnarray}
 2 M(\mathbf{k})\left \{  \frac{1}{g_P}- \frac{N}{yk_BT} \frac{\sinh y}{\cosh y + \cosh \frac{\mu}{k_B T}}\right \} =0,
 \label{mu0}
\end{eqnarray}
where
\begin{eqnarray}
y =\frac{\sqrt{\epsilon(\mathbf{k})^2 + M(\mathbf{k})^2}}{k_BT}.
\label{mu2}
\end{eqnarray}

For $M \neq 0$, namely, inside the PG region, we conclude, from (\ref{mu0})
 that $y$ must satisfy
\begin{eqnarray}
\alpha (x,T)y = \tanh y,
\end{eqnarray}
where
\begin{eqnarray}
\alpha (x,T) = \frac{k_BT}{Ng_P }\frac{\left[ \cosh y + \cosh \left (\frac{\mu(x)}{k_BT}\right) \right]}{\cosh y}.
\label{co2}
\end{eqnarray}
For $\alpha > 1 $ ,the only solution for the equation above is $y_0=0$. Consequently, a Fermi surface does not form in this case, because, according to (\ref{AB}), the Fermi pockets are ellipses with vanishing semi-axes. For $\alpha < 1$, on the other hand, a Fermi surface will appear, with elliptical pockets of finite semi-axes.

The threshold for this to happen, therefore, is $y_0 =0$. The boundary of the region of the phase diagram, where a Fermi surface starts to be seen (Fermi pockets) is, then given by $T_{co}(x)$.
\begin{eqnarray}
 \frac{k_BT_{co}}{Ng_P }=\frac{1}{\left[ 1 + \cosh \left (\frac{\mu(x)}{T_{co}}\right) \right]}.
\label{co}
\end{eqnarray}

We refer the reader to \cite{M6} and \cite{M1} for further reading, especially about the relation between Fermi surface formation and the occurrence of a Charge Density Wave ordered phase in the pseudogap region.


    \section{ 4) The Resistivity of High-Tc Cuprates}
    
We describe here how we can derive from our theory a general expression for the resistivity of the high-Tc cuprates, as well as the effects of an applied magnetic field on it \cite{M2,M3}.

From the very outset, let us remark that, assuming the resistivity is produced by hole-exciton scattering, using a Drude formula approach we must have the resistivity proportional to the exciton density. Since these are bosons, we must have their density regulated by the Bose-Einsten distribution, hence

\begin{eqnarray}
\rho \propto n_{exciton} \sim \frac{1}{e^{\frac{E-\mu}{k_B T}}-1} \sim 
\frac{k_B}{E-\mu}T.
\end{eqnarray}

We easily obtain, thereby, a physical justification to the linear-in-T behavior for the resistivity of the cuprates without the need of any further assumptions.

This is, however, a qualitative physical argument, just pointing a direction.

    \subsection{  4.1) The Resistivity from the Kubo Formula }

In order to obtain an accurate expression for the resistivity of cuprates, we start from the Kubo formula for the conductivity at a finite temperature
\cite{gm}
\begin{eqnarray}
\sigma^{ij}_{\text{DC}}=\lim\limits_{\omega\rightarrow 0}\frac{i}{ \omega}\left[1- e^{-\beta\hbar\omega} \right ]\lim\limits_{\mathbf{k}\rightarrow \mathbf{0}} \Pi^{ij}\left(\omega + i\delta, \mathbf{k}\right),
\label{10z}
\end{eqnarray}
where $\Pi^{ij}$ is the retarded, connected current-current correlation function:
\begin{eqnarray}
\Pi^{ij}=\langle j^{i}j^{j}\rangle_{\text{C}}.         
\end{eqnarray}
This is given by
\begin{eqnarray}
\langle j^i j^j\rangle_C \left(\omega, \mathbf{k}\right) = \frac{\delta^2  \Omega [ \textbf{A} ] }{\delta \textbf{A}^i\left(\omega, \mathbf{k}\right) \delta \textbf{A}^j\left(\omega, \mathbf{k}\right)}\Huge |_{\textbf{A}},
\label{33}
\end{eqnarray}
where $\Omega [ \textbf{A} ]$ is the thermodynamic potential in the presence of an applied electromagnetic vector potential $ \textbf{A}\left(\omega, \mathbf{k}\right)$
\begin{eqnarray}
\Omega [ \textbf{A} ]=-\frac{1}{\beta} \ln Z[\textbf{A}]\ \ \ \ \ ;  \ \ \ \
Z[\textbf{A}]= {\rm Tr}_{Total} e^{-\beta \left[ H[\textbf{A}]-\mu\mathcal{N}\right ]}.
 \label{3}
\end{eqnarray}

In the absence of an applied magnetic field we must consider expression (\ref{33}) at
$\textbf{A}=0$, while in the presence of an external magnetic field, $\textbf{B}=\mu_0 \textbf{H}$, we must take it at $\textbf{A} =\frac{1}{2} \textbf{r}\times \textbf{B}$.

	The expression above follows from the usual minimal coupling prescription
\begin{eqnarray}
\epsilon(\hbar\textbf{k}) \longrightarrow \epsilon(\hbar\textbf{k} +
 e \textbf{A}),
\label{4}
\end{eqnarray}
introduced in the Hamiltonian,
after which, the eigenvalues of $H-\mu\mathcal{N}$ become 
\begin{eqnarray}
  \mathcal{E}_{\pm}^2[\textbf A]=\Delta^2+\Big (\sqrt{v^2 (\hbar\textbf k +e \textbf A)^2 + M^2} \pm\mu \Big )^2.
\label{9}
\end{eqnarray}

In the SC phase, we have $\Delta \neq 0$, $M =0$, and it can be shown \cite{M3} that indeed the resistivity vanishes as it should,

	\begin{eqnarray}
	\rho_{SC}^{ij} 
	 &=&\frac{\delta^{ij} M}{\hbar\beta V^{-1} e^2 v^2  \left \{ \frac{2|\mu|}{\sqrt{\Delta^2+\mu^2}}\tanh\left [\frac{\sqrt{\Delta^2+ \mu^2}}{2k_BT}  \right ] \right \}} \stackrel{M\rightarrow 0}{\longrightarrow 0}.
	\end{eqnarray}

In the non-SC phases, where $\Delta = 0$, conversely, we obtain
with the help of the eigenvalues above, the corresponding general DC resistivity per $CuO_2$ plane, \cite{M3}

\begin{eqnarray}
 \rho = \frac{Vk_B }{\hbar e^2 v^2 } \frac{\mathcal{M} T}{ \left[ \tanh\Big (\frac{ \mathcal{M} + \mu +  \hbar\omega_0}{2k_BT}\Big )+
  \tanh\Big (\frac{ \mathcal{M} + \mu -  \hbar\omega_0}{2k_BT}\Big )+ \tanh\Big (\frac{ \mathcal{M} - \mu +  \hbar\omega_0}{2k_BT}\Big )+
  \tanh\Big (\frac{ \mathcal{M} - \mu -  \hbar\omega_0}{2k_BT}\Big )\right ]},
  \label{ro2}
  	\end{eqnarray}
or, equivalently \cite{M3}
 \begin{eqnarray}
 \rho&=&\frac{Vk_B }{\hbar e^2 v^2 } \frac{\mathcal{M} T}{ \left[\frac{\sinh\left (\frac{ \mathcal{M}  }{k_BT}\right )}{\cosh\left (\frac{ \mathcal{M} }{k_BT}\right ) + \cosh\Big (\frac{ \mu+ \hbar\omega_0}{k_BT}\Big )}+
 \frac{\sinh\left (\frac{ \mathcal{M}  }{k_BT}\right )}{\cosh\left (\frac{ \mathcal{M}  }{k_BT}\right ) + \cosh\Big (\frac{ \mu - \hbar\omega_0}{k_BT}\Big )}\right] } , 
 \label{ro1}
 \end{eqnarray}
   where
   \begin{equation}
\frac{\mathcal{M}}{k_BT} = \sqrt{\left( \frac{M}{k_BT}\right )^2+ 
 \lambda_2^2 \left( \frac{\mu_B\mu_0 H}{k_BT}\right )^2},
\end{equation}
in the presence of an applied magnetic field $ \textbf{B}=\mu_0 \textbf{H}$.
 $\mu_B=\frac{e\hbar}{2m_e}$ is the Bohr magneton, $\frac{\mu_B}{k_B}=0.671 \ K/T$, $\hbar \omega_0=\frac{1}{2}\mu_B\mu_0 H$ is the Zeeman coupling  and $ \lambda_2 \simeq   \frac{m_e}{m}$, where $m$ is the characteristic quasi-particle mass.

In the expressions above $V=da^2$ is the volume of the primitive unit cell, per $CuO_2$ plane, with $d$ being the distance between planes, $a$ the lattice parameter and $N$ is the number of 
$CuO_2$ layers per primitive unit cell. Finally, $v$, is the characteristic velocity of the holes, such that, for LSCO, $\left(\hbar v/a\right)\approx 2.86\times 10^{-2} eV$~\cite{M1}.

We can re-write (\ref{ro2}) and (\ref{ro1}) 
in terms of a three-variable scaling function $G(K_1, K_2, K_3)$, where 
\begin{equation}
    K_1=\frac{M}{k_BT}\ \ ;\ \ K_2=\frac{\mu}{k_BT}
    \ \ ;\ \ K_3=\frac{\mu_B\mu_0 H}{k_BT},
\end{equation}
namely,
\begin{equation}
		\rho(x,T)=BT^2 G\left(\frac{M}{k_B T}, \frac{\mu}{k_B T},\frac{\mu_B\mu_0 H}{k_B T}\right),
		\label{eq_rho}
	\end{equation}

	In the previous expressions the (almost universal) constant $B$ is given by
	\begin{equation}
		B=\frac{h}{e^2}\frac{d}{2\pi}\left(\frac{a}{\hbar v}\right)^2k_B^2\approx 0.37\times d  \  n\Omega \text{cm}/K^2,
		\label{eq_B}
	\end{equation}
	where  $h/e^2\approx 25 812.807 \Omega$ is the von Klitzing constant, namely, the resistance quantum and
 $d$ is inter-plane distance given in \AA\ -units.

    Expression (\ref{eq_rho}) is a general formula for the resistivity of cuprates, valid
    either in the presence of an applied magnetic field or not. 

    We can assert, in general, that resistance in the hole-doped cuprates is a consequence of hole-exciton scattering, which ultimately produces an effective hole-hole scattering. Dissipation, by its turn, is generated from the hole-exciton interaction, mainly by the mechanisms: a) recombination [ exciton + hole1 $\rightarrow$ hole2 + phonons ]; b) exciton-hole scattering with a finite exciton lifetime $\tau$; in which
      an energy $\Delta E = h/\tau$ will be dissipated.

    In the next subsection, we shall explore applications of our theory for describing the resistivity in different phases of the cuprates.

\subsection{ 4.2 The Resistivity in each Phase}

 The peculiar form of the resistivity in each of the different phases will be determined by the form the function $G\left(K_1,K_2,K_3\right)$ assumes in each phase.
	
The scaling function, in the different non-SC phases of cuprates \cite{M2} will determine the specific form of the resistivity in each phase.

In the FL phase, we have $K_1=K_2=K_3=0$ a $T^2$ and the resistivity in this phase,
is given by
\begin{equation}
\rho_{FL}(T)= BT^2,
\end{equation}

with a coefficient $B\simeq2.45\ n\Omega cm/K^2$, which remarkably agrees with the experimental result \cite{fliq} $B\simeq2.50\pm 0.1\ n\Omega cm/K^2$.\cite{M2,M3}.

 Particularly interesting is the strange metal phase, where we have, both the SC and PG parameters vanishing: $\Delta=0$ and $M=0$. 
The chemical potential, conversely, scales with $T$, namely $\mu=D T$, where $D=2.69 \ eV/K$ \cite{M3}. 
Consequently, we will have
$K_1=0$, $K_2= D/k_B$, $K_3=\frac{\mu_B\mu_0 H}{k_BT}$.

The general expression (\ref{eq_rho}) implies the resistivity has a linear dependence on $T$ with a slope proportional to $T^*$ \cite{M2,M3}

\begin{equation}
\rho_{SM}(T)= BCT^* T
\end{equation}
where $C=\cosh\left(\frac{D}{2}\right)$ and $K_2=D$.

\begin{figure}
	[h]
	\centerline
	{
		\includegraphics[scale=0.4]{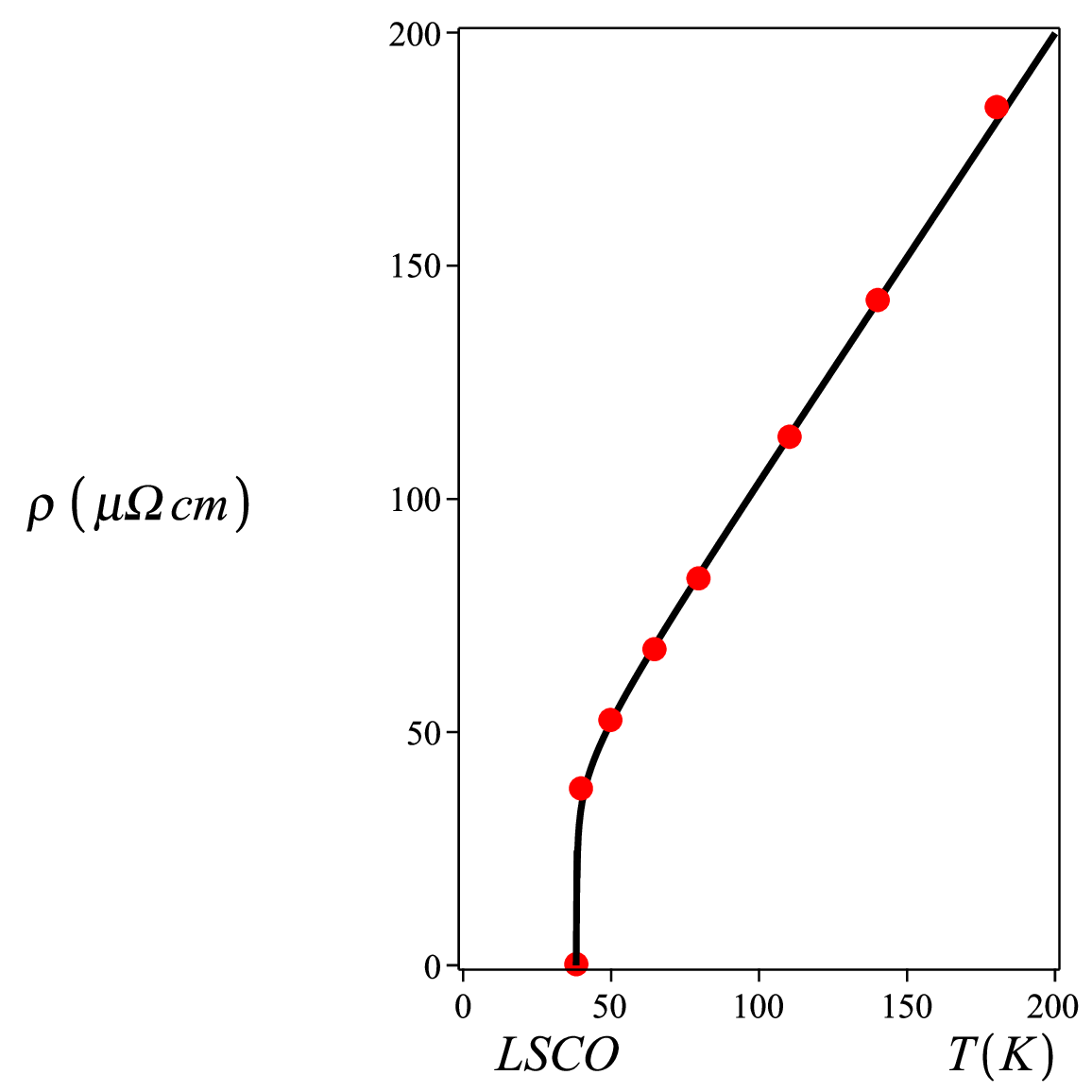}}
		\caption{Resistivity of LSCO at zero magnetic field: $\mu_0H=0$. The solid line is the plot of the theoretical expression derived from our theory, Eq. (\ref{ro1}) for a sample with $T_c=38.5K$ at zero magnetic field. Experimental data from \cite{gg}.}
	\label{f1}
\end{figure}

\begin{figure}
	[h]
	\centerline
	{
		\includegraphics[scale=0.4]{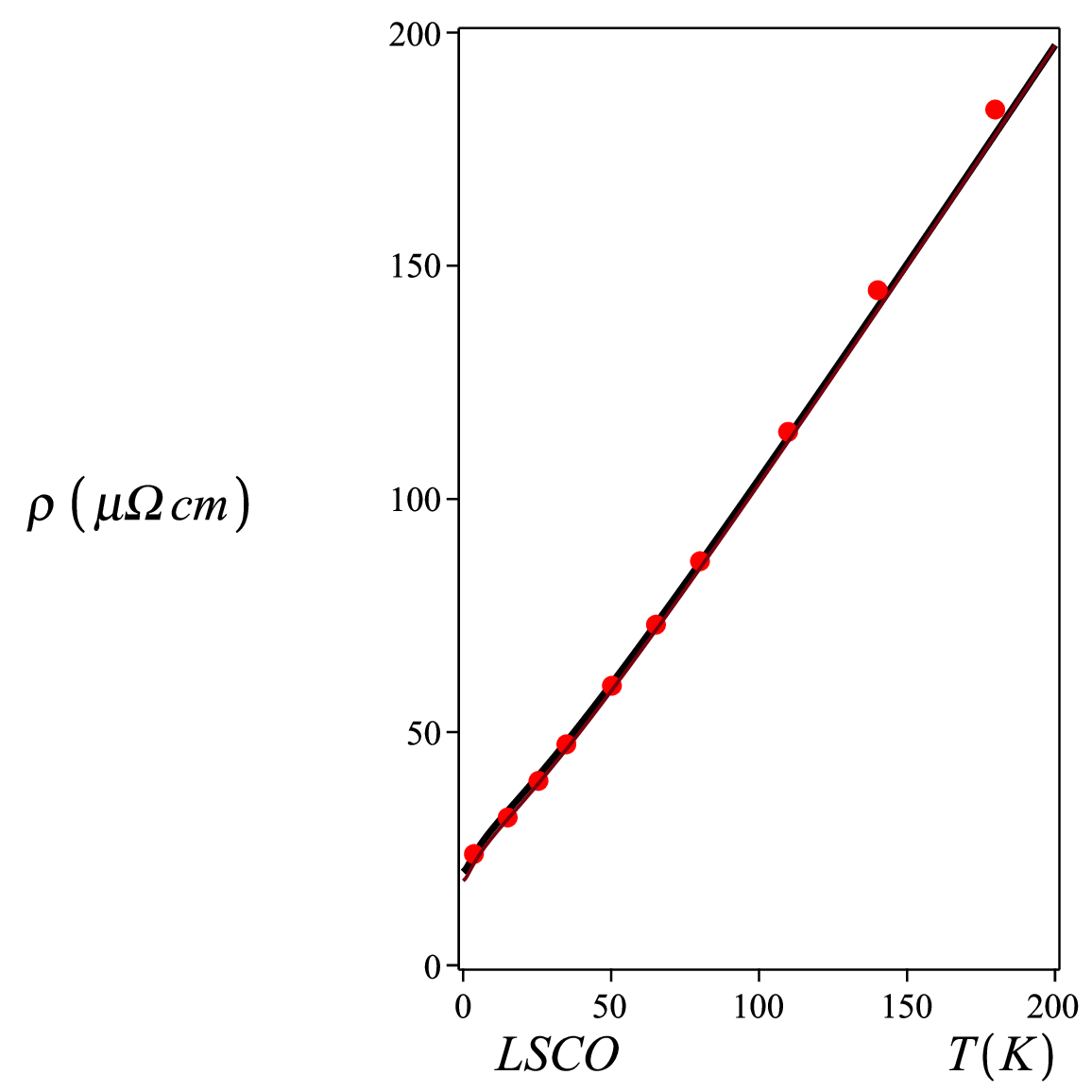}}
	\caption{Resistivity of LSCO at a magnetic field: $\mu_0H=50 T$. The solid line is the plot of the theoretical expression derived from our theory, Eq. (\ref{ro1}) for a sample with $T_c=38.5K$. Experimental data from \cite{gg}.}
	\label{f2a}
\end{figure}

\begin{figure}
	[h]
	\centerline
	{
		\includegraphics[scale=0.4]{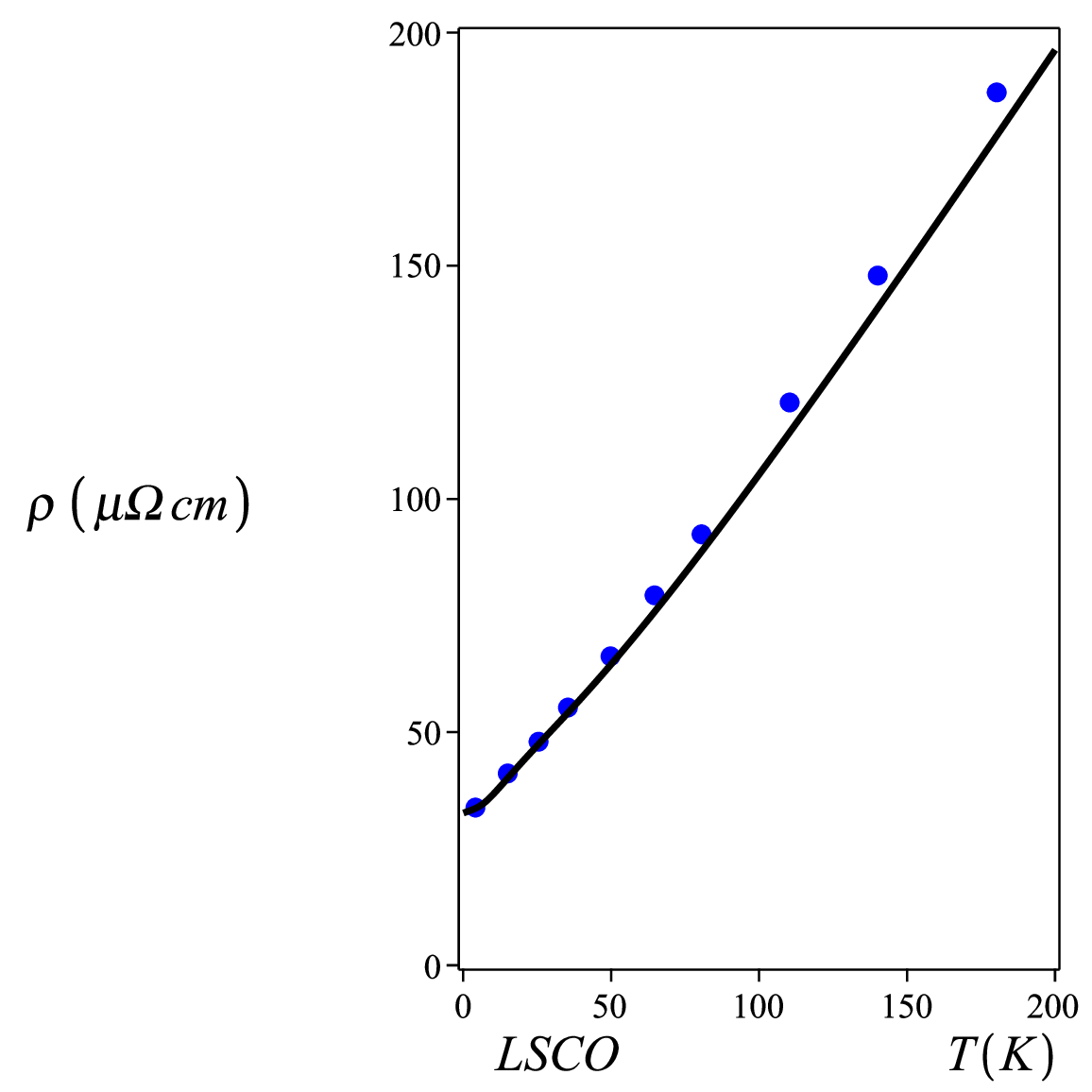}}
		\caption{Resistivity of LSCO at a magnetic field: $\mu_0H=80 T$. The solid line is the plot of the theoretical expression derived from our theory, Eq. (\ref{ro1}) for a sample with $T_c=38.5K$. Experimental data from \cite{gg}.}
	\label{f2}
\end{figure}

In Fig. \ref{f1}  we plot our expression (\ref{ro1}), for the zero field resistivity (solid black line), together with the experimental data from \cite{gg}. In Fig. \ref{f2a}, we represent the curves corresponding to our expression (\ref{ro1}), for an applied magnetic field of $50 T$  with the experimental data from \cite{gg}. Finally, in Fig. \ref{f2}, we display the curve corresponding to (\ref{ro1}) for an applied magnetic field of $80 T$.

In the PG phase, where $K_1\neq 0,K_2\neq 0,K_3=0$, conversely, the resistivity presents an exponential dependence on $T$ for $K_2>K_1$, a quadratic dependence on $T$, for $K_2\approx K_1$ and a linear dependence on $T$ for $K_1 \rightarrow 0$, namely,
\begin{equation}
\rho_{PG}(T) \propto
		\begin{cases}
		T^2 e^{1/T}	\ \ \ \ \  K_2 > K_1 \\
              T^2		\ \ \ \ \  K_2 \simeq K_1  \\
              T  \ \ \ \ \  K_1 \rightarrow 0
  \label{gkk1}
		\end{cases}.
	\end{equation}

We see that our general expression for $\rho(T,H)$ is in agreement with the experimental data for the different phases of LSCO either in the presence of a magnetic field or not.

 \section{ 5) The Effect of an External Applied Pressure on the Phase Diagram}

\subsection{ 5.1) Variation of $T_{max}$  with the External Pressure}

It is natural to expect that the overlap integrals $J_{AF},J_{AF},U,t$, will be
    modified under the action of an external pressure. This should, in principle, influence  the coupling parameters $g_S=\frac{J^2_K}{8J_{AF}}$
    and $g_P=\frac{2t^2}{U}$. However, since $U$ and $J$ possess four overlap integrals each, whereas $t$ possesses just two, it follows that $g_P$ should not change by the action of an external pressure. Consequently, the pseudogap temperature $T^*$, which depends solely on $g_P$, should not be altered, as well, by the application of an external pressure. This is a prediction of our theory. 
    
    The critical temperature $T_c(x)$, however should change because
    it turns out that \cite{M1}
     under a pressure variation $\Delta P$, the magnetic exchange coupling parameters behave as follows
\begin{eqnarray}
\frac{J(P)- J(P_0) }{J(P_0)}=  \kappa_J \Delta P,
\label{p1a}
\end{eqnarray}
where $\kappa_{J}>0$ is the effective compressibility modulus for $J$. 

For an infinitesimal variation of pressure, this can be written as
\begin{eqnarray}
\frac{1}{J}\frac{dJ(P)}{dP}= \kappa_J.
\label{p6}
\end{eqnarray}
whose solution is
\begin{eqnarray}
J(P)=J(0) e^{ \kappa_{J} P}.
\label{ap7}
\end{eqnarray}

Assuming the moduli of compressibility, $\kappa_J$'s, for  the different magnetic couplings, $J_{AF}$, $J_K$,  are approximately the same, we come to the conclusion, that the SC coupling parameter $g_S$ grows exponentially with the pressure, with an effective modulus of compressibility, $\kappa_g$, which must be determined:
\begin{eqnarray}
g_S(P)=g_S e^{ \kappa_{g} P}.
\label{p8}
\end{eqnarray}

\begin{figure}
	[h]
	\centerline
	{
		\includegraphics[scale=0.4]{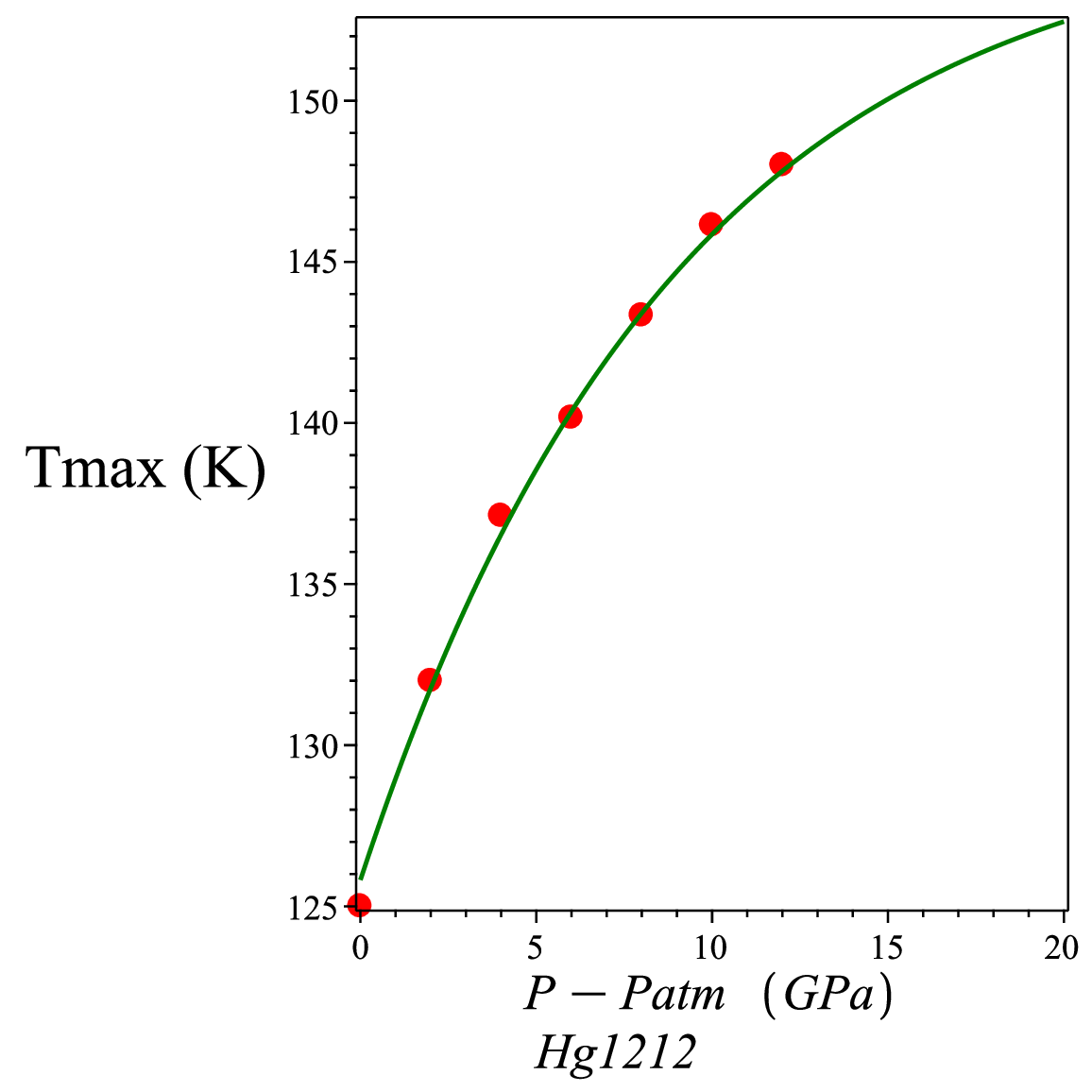}}
		\caption{Effects of an applied external pressure of up to $15 GPa$ in Hg1212. Experimental data from \cite{p}.}
	\label{g222}
\end{figure}

Then, we investigate how the optimal temperature $T_{max}$ evolves with the pressure.
From  (\ref{eqtc}), we see that the optimal SC transition temperature depends on pressure through the function $\eta(Ng_S(p))$. Then, inserting (\ref{p8}) into (\ref{eqtc}),  we obtain the curve for $T_{max}(P)$,for $Hg1212$, depicted in Fig. \ref{g222} after adjusting the parameter $\kappa_g$ to the single value $\kappa_{g}=\frac{1}{17}GPa^{-1}$.

\section{ 6) Conclusion}

Two roads emerge from the simplifications that were obtained from the 
3BHM, respectively associated to the t-J Model, and to the Spin-Fermion-Hubbard Model (SFHM).  The former provides the framework for Anderson's RVB-Theory. In the framework of this theory, an effective ground-state $|GS\rangle$, the RVB-state, was conjectured and it was shown to correctly describe the SC phase of the cuprates.  
Most of the researchers followed this path.

We took a different way, in the attempt to describe High-Tc superconductivity in cuprates, which makes use of the Spin-Fermion-Hubbard Model.
One of the important procedures that was responsible for that  was the very careful choice of the microscopic constituents, which would most appropriate to represent the real system. These perfectly fit the description provided by the Spin-Fermion-Hubbard Hamiltonian.

In a recent publication \cite{M6}, I have shown that the effective Hamiltonian of the holes in the cuprates, derived from the SFHM, possesses as the ground-state, an RVB-state very similar to the one proposed by Anderson. Our RVB state, however is only possible in the presence of the oxygen bipartite lattice because the resonating dimers are Cooper pairs formed by holes belonging, each of them, to nearest points belonging to different sub-lattices A and B. These lattices are no longer present in Anderson's approach. 

The determination of the associated eigenvalues of the relevant Hamiltonian allowed for the obtainment of the thermodynamic potential of the system. This revealed the properties of several physical observables, that could be directly compared with the results of many experiments. This feature made our theory testable. 

The tests to which it has been submitted so far include, among others: the full phase diagram, resistivity, magneto-resistivity, the effects of pressure and of the number of planes on the phase diagram and the depletion of the density of states around the Fermi surfaces seen in the pseudogap phase. \cite{M1,M2,M3,M4,M5,M6}.

It seems clear that if we just guess what is a ground state of the system, thereby explaining the onset of a SC state, without any further knowledge about the Hamiltonian, partition function and thermodynamic potentials, then just a few things can be said about such a system. This procedure, which was sufficient for explaining the Quantum Hall Effect is clearly not sufficient to describe the whole complexity of the cuprates. 

Our theory, despite exhibiting a very similar ground state in the SC phase, does not suffer from these limitations and therefore has been able to reproduce a large amount of experimental measurements performed in the cuprates. It has also made several predictions that can be tested. Just to mention one of these: it predicts that the pseudogap temperature $T^*(x)$ will not be modified by the application of an external hydrostatic pressure \cite{M1,M2,M3,M4,M5,M6}. 

Another important prediction that emerges from our theory for hole doped cuprates is that the mechanism responsible for SC in the electron doped cuprates must be a different one. It is likely that further study along the lines we described here may lead to the comprehension of SC in these materials in the near future.

The formulation of our theory for  High-Tc SC
in hole doped cuprates was preceded by a preparatory work, where some ideas like the competition between the SC phase and the
exciton condensate phase and the Spin-Fermion-Hubbard model were tested in systems with Dirac electrons \cite{MM,MM1,MM2,MM3,MM4,MM5}.
Also the foundation of the approach used for describing the spin-glass phase in cuprates\cite{M6} were laid down in \cite{sg1,sg2,sg3}. 

\section{Appendix: From the Three Bands Hubbard Model to the Spin-Fermion-Hubbard Model  }

In this Appendix, we review the derivation of the Heisenberg and Kondo terms which appear in the Spin-Fermion-Hubbard Model. This can be found in the literature (see \cite{dk}, for instance) and is included for the reader's convenience.

\subsection{A.1) Direct Exchange: from the $d$-Band Hubbard Model to the Heisenberg Model}

Given a system with direct exchange between neighbor spins, which results from the hybridization of the relevant orbitals, namely, the direct overlap of neighbor atomic wave-functions, we can write the corresponding Hubbard Hamiltonian as

 \begin{eqnarray}
 &\ &
H = H_0 +H_1
\nonumber \\
&\ &H_0 = E_d \sum_{I\sigma} n_{I\sigma}^d + U_d \sum_I n^d_\uparrow n^d_\downarrow
\nonumber \\
&\ &H_1=-t \sum_{\langle IJ\rangle}\sum_\sigma\left[ d^\dagger_{I\sigma}d_{J\sigma}
+ d^\dagger_{J\sigma}d_{I\sigma}\right ].
\label{hubb}
\end{eqnarray}
Treating the last term above in 2nd order Rayleigh-Schr\"odinger pertubation theory
\cite{ecm2}, we obtain 
\begin{equation}
H_{eff} = \sum_{n\neq 0} \frac{H_1 |n\rangle\langle n|H_1}{E_n^{(0)}-E_0^{(0)}},
 \end{equation}
where the sum sweeps the relevant intermediate states.
Considering that the energy of these intermediate states, at half-filling is given by
\begin{equation}
E_0^{(0)}- E_1^{(0)}=U_d 
\label{xxx},
\end{equation}

we have
\begin{eqnarray}
&\ & H_{eff} = -\frac{1}{U_d}H_1 H_1
\nonumber \\
&\ & =\frac{2t^2}{U_d}\sum_{\langle IJ\rangle}\sum_{\alpha,\beta}\left [d^\dagger_{I\alpha} d_{I\beta} d^\dagger_{J\lambda} d_{J\rho}\right ]\delta_{\alpha\rho} \delta_{\beta\lambda}
 \end{eqnarray}
Here we used
the identity
$$
\vec{\sigma }_{\alpha\beta}\cdot \vec{\sigma }_{\rho\lambda}=2
\delta_{\alpha\lambda} \delta_{\beta\rho}.
$$

Now, considering
 that the spin operator for an electron at site $I$ is given by
\begin{equation}
\mathbf{S_I}= \frac{1}{2}d^\dagger_{I\alpha} \vec{\sigma }_{\alpha\beta}d_{I\beta}
\label{xxx1}
\end{equation}

we arrive at the conclusion that the effective interaction which emerges from the Hubbard model, Eq. (\ref{hubb}), is described by
the AF Heisenberg Hamiltonian , namely,
\begin{equation}
H_{eff} = \frac{4t^2}{U_d}\sum_{\langle IJ\rangle}
 \textbf{S}_I\cdot  \textbf{S}_J
\end{equation}
in which 
the effective AF coupling is
$J_{AF}=\frac{4t^2}{U_d}$.

\subsection{A.2) Super-Exchange: The Effective IJ Hopping}

 Now, the Hamiltonian, describing the relevant interaction reads
 \begin{eqnarray}
H = E_d \sum_{I\sigma} n_{I\sigma}^d + E_p \sum_{R\in \{R_A\},\{R_B\};\sigma} n_{R;\sigma}^p + U_{pd} \sum_I n^d_\uparrow n^p_\downarrow
-t_{pd} \sum_{\langle IJ\rangle}\sum_\sigma\left\{\left[ d^\dagger_{I\sigma}p_{A\sigma}
+ p^\dagger_{A\sigma}d_{I\sigma}\right ] +\left[ d^\dagger_{J\sigma}p_{A\sigma}
+ p^\dagger_{A\sigma}d_{J\sigma}\right ]\right\} .
\nonumber \\
+A\leftrightarrow B
\nonumber \\
\label{x1}
\end{eqnarray}
Notice that the direct exchange d-d term is no longer present, thus reflecting the fact that the d-orbitals do not overlap. The d-p overlap, conversely, leads to an effective d-d hopping $\bar t$. Here $p_{A\alpha}, p_{B\alpha}$ are electron creation operators on the oxygen orbitals of $A,B$ sub-lattices.

Intermediate states contributing to the 2nd order Rayleigh-Schr\"odinger perturbation result
are depicted in Figs. \ref{fff8},\ref{ffx8}. We have
\begin{eqnarray}
E_0  - E_1^{(0)} = \Delta E +U_{pd}
\end{eqnarray}
\begin{figure}
	[h]
	\centerline{
		\includegraphics[scale=0.5]{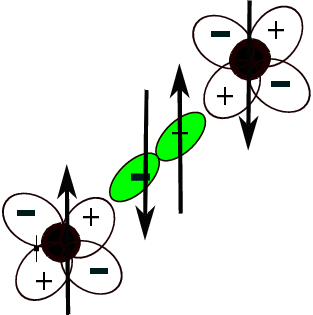}
	}
	\caption{Ground-state of the unperturbed Hamiltonian in Rayleigh-Schr\" odinger perturbation theory. }
	\label{ff8}
\end{figure}

\begin{figure}
	[h]
	\centerline{
		\includegraphics[scale=0.5]{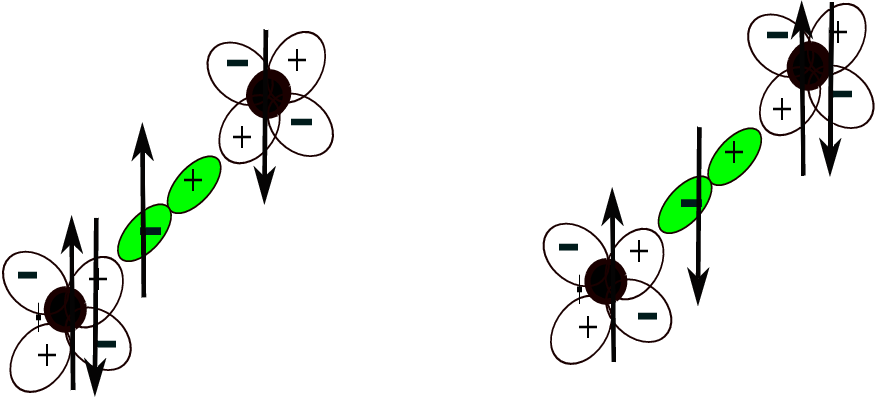}
	}
	\caption{Intermediate states that contribute to the effective Heisenberg Hamiltonian in 2nd. order Rayleigh-Schr\" odinger perturbation theory. }
	\label{fff8}
\end{figure}

\begin{figure}
	[h]
	\centerline{
		\includegraphics[scale=0.5]{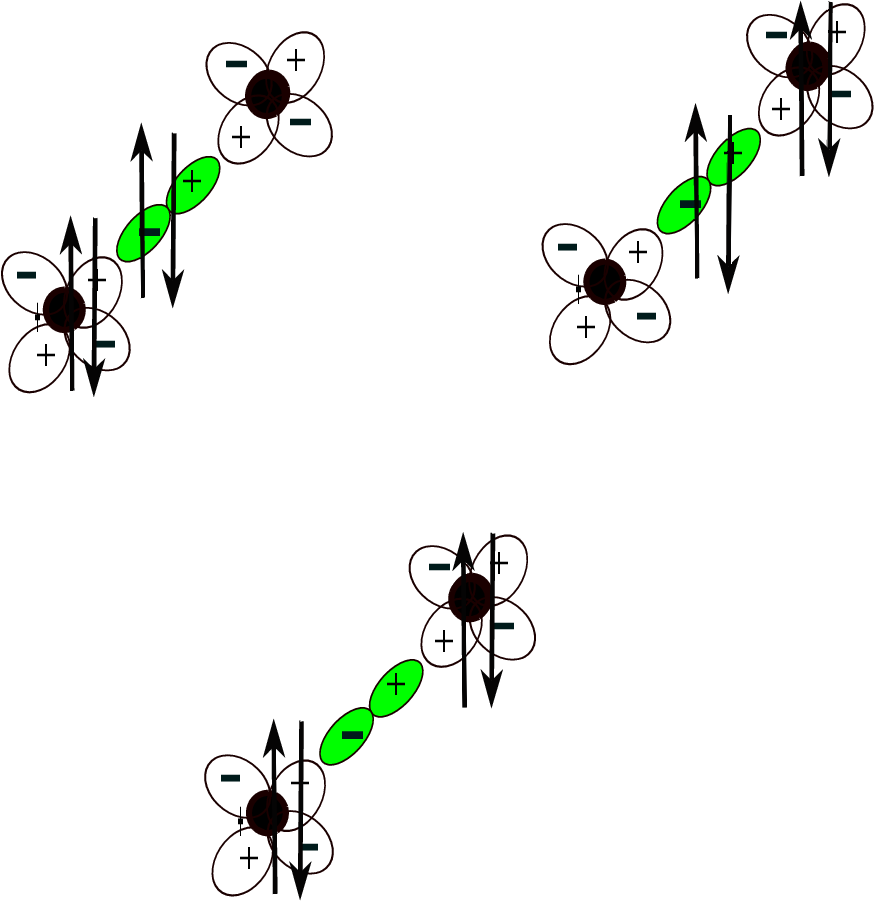}
	}
	\caption{Intermediate states that contribute to the effective  Hamiltonian in 2nd. order Rayleigh-Schr\" odinger perturbation theory. }
	\label{ffx8}
\end{figure}

Using 2nd. order Rayleigh-Schr\" odinger perturbation theory, we find the effective $\langle IJ \rangle$ hopping term

\begin{eqnarray}
H = E_d \sum_{I\sigma} n_{I\sigma}^d + E_p \sum_{R\in \{R_A\},\{R_B\};\sigma} n_{R;\sigma}^p + U_{pd} \sum_I n^d_\uparrow n^p_\downarrow
\nonumber \\
-\left (\frac{2t^2_{pd}}{\Delta E + U_{pd}}\right ) \sum_{\langle IJ\rangle}\sum_{\sigma}\left[ d^\dagger_{I\sigma}d_{J\sigma}
+ d^\dagger_{J\sigma}d_{I\sigma}\right ]
\sum_{\alpha}p^\dagger_{A\alpha} p_{A\alpha}, 
\label{x3}
\end{eqnarray}
where $\Delta E = E_p - E_d$.
The last factor in (\ref{x3}) equals one for the intermediate states of Fig.\ref{fff8}, consequently, we obtain the effective hopping term
\begin{eqnarray}
-\bar t \sum_{\langle IJ\rangle}\sum_{\sigma}\left[ d^\dagger_{I\sigma}d_{J\sigma}
+ d^\dagger_{J\sigma}d_{I\sigma}\right ],
\label{x2}
\end{eqnarray}
where the effective hopping parameter is

\begin{eqnarray}
\bar t=\left (\frac{2t^2_{pd}}{\Delta E + U_{pd}}\right ).
\end{eqnarray}

\subsection{A.3) Super-Exchange: The Effective Heisenberg Hamiltonian}

In terms of the effective hopping parameter above, we can write the relevant three bands Hubbard model Hamiltonian as
\begin{eqnarray}
H &\ &= E_d \sum_{I\sigma} n_{I\sigma}^d + E_p \sum_{R\in \{R_A\},\{R_B\};\sigma} n_{R;\sigma}^p +
\nonumber \\
 &\ & U_{p} \sum_I n^p_\uparrow n^p_\downarrow + U_{d} \sum_I n^d_\uparrow n^d_\downarrow+U_{pd} \sum_I n^d_\uparrow n^p_\downarrow
  \nonumber \\
&\ & -\bar t \sum_{\langle IJ\rangle}\sum_{\sigma}\left[ d^\dagger_{I\sigma}d_{J\sigma}
+ d^\dagger_{J\sigma}d_{I\sigma}\right ].
\label{x44}
\end{eqnarray}

Once again performing 2nd. order Rayleigh-Schr\" odinger perturbation theory
in $\bar t$ and using the intermediate states of Fig. (\ref{ffx8}),for which
\begin{eqnarray}
E^{(0)}_0 - E_1^{(0)} = 2\Delta E +U_{p}
\nonumber \\
E^{(0)}_0 -E_2^{(0)} =  U_{d},
\end{eqnarray}
we obtain the effective Heisenberg interaction Hamiltonian, corresponding to (\ref{x44}), namely,
\begin{equation}
H_{AF} = J_{AF}\sum_{\langle IJ\rangle}
 \textbf{S}_I\cdot  \textbf{S}_J,
\end{equation}
where
\begin{equation}
 J_{AF} = \frac{4t^4_{pd}}{\left (\Delta E + U_{pd} \right )^2}\left [ \frac{1}{U_d}+\frac{2}{2\Delta E +U_p} \right ].
 \label{jaf}
\end{equation}

\begin{figure}
	[h]
	\centerline{
		\includegraphics[scale=0.5]{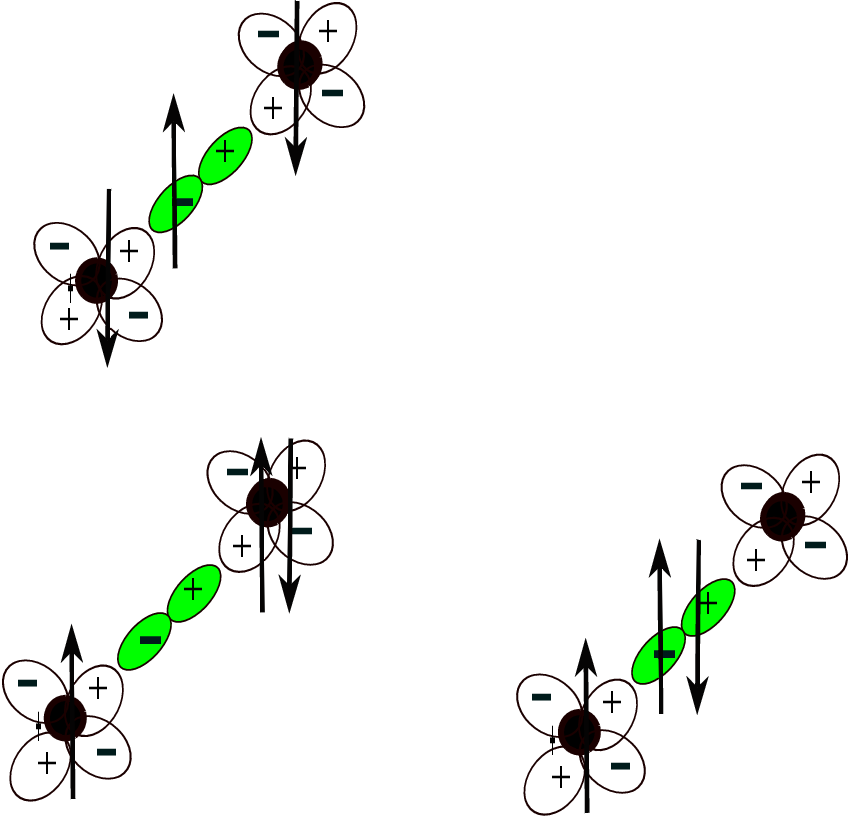}
	}
	\caption{Intermediate states that contribute to the effective  Hamiltonian in 2nd. order Rayleigh-Schr\" odinger perturbation theory in the case that there is one doped hole in the system. }
	\label{ffx9}
\end{figure}

\subsection{A.4) The Effective Kondo Hamiltonian}

Now, when performing 2nd order Rayleigh-Schr\"odinger perturbation theory, the intermediate states that are going to contribute are depicted in Fig. \ref{ffx9}. 
The relevant energy differences to be used become now
\begin{eqnarray}
\epsilon^{(0)}_0 - \epsilon_1^{(0)} = \Delta E -U_{d}
\nonumber \\
\epsilon^{(0)}_0 -\epsilon_2^{(0)} =  \Delta E.
\end{eqnarray}

The effective Kondo electron-hole interaction becomes
\begin{eqnarray}
&\ &
H_{Kondo} = t^2_{pd}\left[\frac{1}{\Delta E }+ \frac{1}{U_d-\Delta E}  \right ]
\sum_{I\alpha\beta}d^\dagger_{I\alpha} d_{I\beta}\left[ \left(\psi^\dagger_{A} +\psi^\dagger_{B}\right )_\alpha  \left( \psi_{A} -\psi_{B} \right )_\beta +
\left(\psi^\dagger_{A} -\psi^\dagger_{B}\right )_\alpha  \left( \psi_{A} +\psi_{B} \right )_\beta
\right ]
\nonumber \\
 &\ & = 
 t^2_{pd}\left[\frac{1}{\Delta E }+ \frac{1}{U_d-\Delta E}  \right ]\sum_{I}\sum_{\alpha\beta}\ d^\dagger_{I\alpha}d_{I\beta}\left(\psi^\dagger_{A\beta}\psi_{A\alpha}- \psi^\dagger_{B\beta}\psi_{B\alpha}\right ).
\label{x611}
\end{eqnarray}

In short, the effective Kondo Hamiltonian describing the interaction between the doped holes, living in the A,B oxygen sublattices, and the localized copper spins is given by 
\begin{eqnarray}
H_K =J_K\sum_{I,\textbf{R}_A\textbf{R}_B} \textbf{S}_I\cdot\mathcal{S},
\label{abc}
\end{eqnarray}
where
\begin{eqnarray}
J_K = t^2_{pd}\left[\frac{1}{\Delta E }+ \frac{1}{U_d-\Delta E}  \right ]
\label{abc1}
 \end{eqnarray}
\begin{eqnarray}
 \mathcal{S} =
\left[\sum_{\textbf{R}_A\in I}   \mathcal{S}_A -
\sum_{\textbf{R}_B\in I}     \mathcal{S}_B   \right ]
\label{abc2}
\end{eqnarray}
The factors $\eta_A,\eta_B,\eta_C,\eta_C'= \pm 1$ correspond to the signs of the lobes of the p, and d orbitals, according to Fig. \ref{fcucell}. In terms of these, we can write

\begin{eqnarray}
H_K =J_K\sum_{I} \textbf{S}_I\cdot\left[\sum_{\textbf{R}_A\in I} \eta_A \eta_C\  \mathcal{S}_A +
\sum_{\textbf{R}_B\in I} \eta_B \eta_C' \    \mathcal{S}_B   \right ]
\end{eqnarray}

\begin{eqnarray}
\textbf{S}_I =\frac{1}{2}d^\dagger_{I\alpha}\vec \sigma_{\alpha\beta}d_{I\beta}
\nonumber \\
  \mathcal{S}_A= \frac{1}{2}\psi_\alpha^\dagger(\textbf{R}_A\in I)\vec \sigma_{\alpha\beta}\psi_\beta(\textbf{R}_A\in I)
  \nonumber \\
  \mathcal{S}_B= \frac{1}{2}\psi_\alpha^\dagger(\textbf{R}_B\in I)\vec \sigma_{\alpha\beta}\psi_\beta(\textbf{R}_B\in I).
  \end{eqnarray}

Now,the specific form of the Ligand orbitals is determined by the symmetry of the corresponding d-orbital, belonging to the $3d^9$ Cu orbital, to which it couples. As we know, this has a $d_{x^2-y^2}$ symmetry, implying the Ligand orbital will be
\begin{eqnarray}
\psi_L=\frac{1}{2}\left [  \psi_{A_1} - \psi_{B_2} - \psi_{A_3} + \psi_{B_4}\right ].
\label{x7}
\end{eqnarray}
In short, the effective Kondo Hamiltonian describing the interaction between the doped holes, living in the A,B oxygen sub-lattices, and the localized copper spins is
given by (\ref{abc}), (\ref{abc1}) and (\ref{abc2}).  

\bigskip
{\bf Acknowledgements}
\bigskip

This work is dedicated to my friend Amir Caldeira, on the occasion of his 75th. birthday. I am grateful to many colleagues whose constructive comments have helped a lot the improvement of this work. Among these I would like to specially mention: Cristiane de Morais Smith, Harry Westphal Jr., Ricardo Doretto, Antonio Helio de Castro Neto and Luiz Nunes de Oliveira. 
I acknowledge the support of the INCT project
Advanced Quantum Materials, involving the Brazilian agencies CNPq (Proc. 408766/2024-7), FAPESP (Proc. 2025/27091-3), and CAPES

\bigskip
{\bf Conflict of Interest}
\bigskip

The author declares that he has no conflicts of interest.

\vfill\eject

\end{document}